\documentclass{jfm}

\usepackage{color}

 \usepackage{graphicx}
\usepackage{natbib}
\usepackage{amsmath}
\ifCUPmtlplainloaded \else
  \checkfont{eurm10}
  \iffontfound
    \IfFileExists{upmath.sty}
      {\typeout{^^JFound AMS Euler Roman fonts on the system,
                   using the 'upmath' package.^^J}%
       \usepackage{upmath}}
      {\typeout{^^JFound AMS Euler Roman fonts on the system, but you
                   dont seem to have the}%
       \typeout{'upmath' package installed. JFM.cls can take advantage
                 of these fonts,^^Jif you use 'upmath' package.^^J}%
       \providecommand\upi{\pi}%
      }
  \else
    \providecommand\upi{\pi}%
  \fi
\fi

\ifCUPmtlplainloaded \else
  \checkfont{msam10}
  \iffontfound
    \IfFileExists{amssymb.sty}
      {\typeout{^^JFound AMS Symbol fonts on the system, using the
                'amssymb' package.^^J}%
       \usepackage{amssymb}%
         \let\leq=\leqslant
         \let\geq=\geqslant
      }{}
  \fi
\fi

\usepackage{hyperref}
\hypersetup{
	hyperindex,
	breaklinks,
	colorlinks=true,
	linkcolor=blue,
	citecolor=blue,
	bookmarks=true,
	bookmarksopen=true,
	bookmarksopenlevel=2,
	pdfstartpage={1},
	pdfstartview={FitH},
	pdfview={FitH 0},
	pdfauthor={Nikolaos A. Bakas and Petros J. Ioannou},
	pdftitle={A theory for the emergence of coherent structures in beta-plane turbulence},
 }
\usepackage[all]{hypcap}	

\ifCUPmtlplainloaded \else
  \IfFileExists{amsbsy.sty}
    {\typeout{^^JFound the 'amsbsy' package on the system, using it.^^J}%
     \usepackage{amsbsy}}
    {\providecommand\boldsymbol[1]{\mbox{\boldmath $##1$}}}
\fi

\newcommand\Real{\mbox{Re}} 
\newcommand\Imag{\mbox{Im}} 

\newsavebox{\astrutbox}
\sbox{\astrutbox}{\rule[-5pt]{0pt}{20pt}}

\title[A theory for the emergence of coherent structures in beta-plane turbulence]{A theory for the emergence of coherent structures in beta-plane turbulence}

\author[N. A. Bakas and P. J. Ioannou]%
{Nikolaos A. Bakas$^1$%
  \thanks{Email address for correspondence: nbakas@post.harvard.edu},\ns
and Petros J. Ioannou$^1$}

\affiliation{$^1$National and Kapodistrian University of Athens, Build. IV, office 34,
Panepistimiopolis, Zografos, Athens, Greece\\[\affilskip]}

\pubyear{2012}
\volume{650}
\pagerange{119--126}
\date{?; revised ?; accepted ?. - To be entered by editorial office}
\begin{document}

\maketitle

\begin{abstract}
Planetary turbulent flows are observed to  self-organize into large scale structures such as
zonal jets and coherent vortices. One of the simplest models of planetary turbulence is obtained by
considering a barotropic flow on a beta-plane channel with turbulence sustained by random stirring.
Non-linear integrations of this model show that as the energy input rate of the forcing is increased, the
homogeneity of the flow is broken with  the emergence of non-zonal, coherent, westward
propagating structures and   at larger energy input rates by the emergence of zonal jets.
We study the emergence of non-zonal coherent structures using a non-equilibrium statistical theory,
Stochastic Structural Stability Theory (S3T, previously referred to as SSST). S3T directly models a second order
approximation to the statistical mean turbulent state and allows  identification of statistical turbulent equilibria
and study of their stability. Using S3T, the bifurcation properties of the homogeneous state in barotropic beta-plane
turbulence are determined. Analytic expressions for the zonal and  non-zonal large scale coherent flows that emerge as a result of structural
instability are obtained. Through numerical integrations of the S3T dynamical system, it is found that the unstable structures
equilibrate at finite amplitude. Numerical simulations of the nonlinear equations confirm the
characteristics (scale, amplitude and phase speed) of the structures predicted by S3T.
\end{abstract}

\section{Introduction}

Atmospheric and oceanic turbulence is commonly observed to be organized into spatially and temporally coherent
structures such as zonal jets and coherent vortices. Examples from planetary turbulence include the
banded jets and the Great Red Spot in the Jovian atmosphere \citep{Ingersoll-90,Vasavada-05},
as well as the latent jets in the Earth's ocean basins \citep{Maximenko-05} and the ocean rings shed by the
meandering of the Gulf-Stream in the western Atlantic Ocean \citep{Chelton-07}. Laboratory experiments and
numerical simulations of both decaying and forced turbulence have shown that these coherent structures
appear and persist for a very long time despite the presence of eddy mixing \citep{Vallis-93,Cho-96,Weeks-etal-97,Read-etal-2004,Espa-etal-2010,DiNitto-etal-2013}.

One of the simplest models of planetary turbulence, is the
stochastically forced barotropic vorticity equation on the surface of a
rotating planet or on a $\beta$-plane (a plane tangent to the surface of the planet in which differential
rotation is taken into account). A large number of numerical simulations of this model have shown that robust,
large scale zonal jets emerge in the flow and are sustained at finite amplitude \citep{Williams-78,Vallis-93,
Nadiga-06,Danilov-04,Galperin-etal-06}. In addition, large scale westward propagating coherent waves were found
to coexist with the zonal jets \citep{Sukoriansky-etal-2008,Galperin-etal-10}. These waves were found to either
obey a Rossby wave dispersion, or propagate with different phase speeds. The propagating
waves typically have low zonal wavenumbers and were found in a parameter regime in which strong, robust jets
dominate. These waves that are referred to as satellite modes \citep{Danilov-04} or zonons \citep{Sukoriansky-etal-2008},
appear to be sustained by non-linear interactions between Rossby waves. However the mechanism for their excitation and
maintenance remains elusive. The goal in this work is to develop a non-equilibrium statistical theory that can
predict the emergence of both zonal jets and non-zonal coherent structures and can capture their characteristics.

The tendency for formation of large scale structures in planetary turbulence can be understood in terms of the
approximate energy and vorticity conservation in two dimensional or quasi two dimensional flows
that implies an energy transfer from small to large scales given that there is a direct enstrophy cascade to
small scales \citep{Fjortoft-1953}. \cite{Rhines-75} found that the
non-linear eddy-eddy interactions that are local in wavenumber space, lead to an inverse energy cascade that
is anisotropic, as it is inhibited in the region in wavenumber space in which weakly interacting Rossby-waves dominate.
The cascade therefore continues through a narrow region in wavenumber space, transferring energy to
zonal jets \citep{Vallis-93,Nazarenko-09} and is finally arrested at a meridional scale that is dictated by
friction \citep{Smith-etal-02,Sukoriansky-etal-2007}. However, observations of the atmospheric
midlatitude jet \citep{Shepherd-87} and numerical analysis of simulations
\citep{Nozawa-97,Huang-98,Huang-01}, showed that the large scale jets are maintained through
spectrally non-local interactions rather than by a local in wavenumber space cascade. Theoretical
studies \citep{FI-03,FI-07} and numerical simulations
\citep{Srinivasan-Young-12,Constantinou-etal-2012} have also shown that jets emerge even in the absence of cascades.
Moreover, the
persistence and dominance of specific non-zonal coherent structures that also emerge cannot be explained by the
phenomenological description of the inverse turbulent cascade.

Since organization of turbulence into coherent structures involves complex non-linear interactions among a large number
of degrees of freedom, an alternative approach for gaining an understanding for the tendency towards self-organization
of turbulent flows is to use statistical mechanics, an approach pioneered by \cite{Miller-1990} and
\cite{Robert-Sommeria-91} in what is now known as Robert-Sommeria-Miller (RSM) theory. The RSM theory builds upon the
work of \cite{Onsager-1949} that explains self-organization of turbulence in terms of the equilibrium statistical
mechanics of a set of point vortices. The main idea is to find a solution of
the unforced Euler equations that maximizes a proper measure of entropy under the restrictions imposed by
all conserved quantities. The coherent structures that emerge from this statistical analysis for
two dimensional and quasi-geostrophic flows are either large scale vortices \citep{Chavanis-Sommeria-1998} or jets
\citep{Bouchet-Sommeria-2002,Venaille-Bouchet-2011} (see also a recent review by \cite{Bouchet-2012}). However, the
relevance of these results in planetary flows that are strongly forced and dissipated and are therefore out of equilibrium
remains to be shown.

The emergence of coherent structures in barotropic turbulence has also another feature that needs to be
explained. As the energy input of the stochastic forcing is increased or the dissipation is decreased,
nonlinear simulations show that there is a sudden emergence of coherent zonal flows
\citep{Srinivasan-Young-12,Constantinou-etal-2012} and as will
be shown in this work of non-zonal coherent structures as well. This argues that the emergence of coherent
structures in a homogeneous background of turbulence is a bifurcation phenomenon, as is for example the
formation of patterns in thermal convection. In this case, Rayleigh's theory
of hydrodynamic instability \citep{Rayleigh-16} and the extension of the theory to the weakly nonlinear and
fully nonlinear regime was able to predict the critical Rayleigh number for the onset of the convective regime as
well as the scales and amplitude of the emerging structures \citep{Busse-78}. The emergent structures take the form
among others of stationary striped patterns and oscillating cells \citep{Cross-Greenside-2009} which are like the zonal jets
and the westward propagating structures that emerge in barotropic beta-plane turbulence
\citep{Parker-Krommes-2013,Bakas-Ioannou-2013}. The difficulty in obtaining
such a stability theory in the case of planetary flows, is that in contrast to thermal convection, the basic state is
a complex time--dependent solution of the Navier-–Stokes equations and not a stationary point of the equations.

An alternative approach is to study the dynamics and stability of the statistical equilibria, which are fixed points of the equations
governing the evolution of the flow statistics. This approach is followed
in the Stochastic Structural Stability
Theory (S3T previously referred to as SSST) \citep{FI-03} or
Second Order Cumulant Expansion theory (CE2) \citep{Marston-etal-2008}, which is a non-equilibrium statistical theory that was
applied to macroscale barotropic and
baroclinic turbulence in
planetary atmospheres, wall bounded turbulence, plasmas and astrophysical flows \citep{FI-03,FI-07,FI-08,Marston-etal-2008,Farrell-Ioannou-2009-equatorial,Farrell-Ioannou-2009-plasmas,Farrell-Ioannou-2009-closure,
Marston-2010,Tobias-etal-2011,Srinivasan-Young-12,Marston-2012,FI-12}. This theory is based on two building blocks. The first
is to do a Reynolds decomposition of the dynamical variables into the sum of a mean value that represents the coherent flow
and fluctuations that represent the turbulent eddies and then form the cumulants containing the information on the mean values
(first cumulant) and on the eddy statistics (higher order cumulants). The second building block is to truncate the equations
governing the evolution of the cumulants at second order by either parameterizing the terms involving the third
cumulant \citep{FI-93d,FI-93e,FI-93f,Sole_Farrell-96,DelSole-04} or setting the third cumulant to zero \citep{Marston-etal-2008,Tobias-etal-2011,Srinivasan-Young-12}. Restriction of the dynamics to the first two cumulants is equivalent to
neglecting the eddy-eddy interactions in the fully non-linear dynamics and retaining only the interaction between the eddies
with the instantaneous mean flow. A related approach was also followed by Dubrulle and collaborators
\citep{Dubrulle-Nazarenko-1997,Laval-etal-2003} to describe the interaction of coherent vortical structures with turbulence. While such a second order closure might seem crude at first sight, there is strong evidence
to support it. Previous studies of planetary turbulence have shown that this second order closure produces accurate quadratic
eddy statistics and mean flows \citep{Sole_Farrell-96,DelSole-04,OGorman-Schneider-2007}. In addition, a very recent study that
uses stochastic averaging techniques has shown that for $\beta=0$ and in the limit of weak forcing and dissipation,
the formal asymptotic expansion of the whole probability density function of the non-linear dynamics around a mean flow 
that is assumed to have a singular spectrum of modes, comprises of the
second order S3T closure with an additional stochastic term forcing the mean flow. Therefore S3T accurately describes the
statistical equilibrium mean flow and the eddy statistics, as the additional stochastic term only produces fluctuations
around this statistical equilibrium \citep{Bouchet-etal-2013}.

One of the advantages of S3T is that the nonlinear system governing the evolution of the first two cumulants is
autonomous and deterministic. Its fixed points define statistical equilibria, whose instability brings about
structural reconfiguration of the mean flow and the turbulent statistics. It is therefore amenable to the
usual treatment of classical linear and non-linear stability analysis and actually possesses the mathematical
structure of the dynamical system of pattern formation \citep{Parker-Krommes-2013}. Previous studies employing
S3T have already addressed the bifurcation from a homogeneous turbulent regime to a jet forming regime in barotropic beta-plane
turbulence and identified the emerging jet structures both numerically \citep{FI-07} and analytically
\citep{Bakas-Ioannou-2011,Srinivasan-Young-12} as linearly unstable modes to the homogeneous turbulent state equilibrium.
Comparison of the results of the stability analysis with direct numerical simulations have shown that
the structure of zonal flows that emerge in the non-linear simulations can be predicted by S3T
\citep{Srinivasan-Young-12,Constantinou-etal-2012}. These studies however assumed that the ensemble average is
equivalent to a zonal average, a simplification that treats the non-zonal structures as incoherent and cannot
address their emergence and effect on the jet dynamics.


In order to investigate the dynamics of the coherent non-zonal structures, we adopt in this work the more
general interpretation that the ensemble average represents a Reynolds average with the ensemble mean representing
coarse-graining, an interpretation that has also been recently adopted in S3T studies of baroclinic turbulence
\citep{Bernstein-2009,Bernstein-Farrell-2010}. With this interpretation of the ensemble mean, we obtain the
statistical dynamics of the interaction of both zonal and non-zonal coherent structures with stochastically
forced turbulence on a barotropic $\beta$-plane channel, with the goal of addressing their emergence and
characteristics. We find that the turbulent equilibrium that is homogeneous, is structurally unstable when the
energy input rate is above a threshold and both zonal and non-zonal coherent structures emerge. We also show that the
characteristics of these structures observed in the non-linear simulations are predicted by S3T.

This paper is organized as follows. In section 2 we present the characteristics of the zonal and non-zonal coherent
structures that emerge in non-linear simulations of the turbulent flow. In section 3 we derive the S3T system that
governs the evolution of the ensemble mean coherent structures (first cumulant) and the associated eddy statistics
(second cumulant). In section 4 we analytically study the instability of the corresponding homogeneous equilibrium,
analyzing the unstable structures and their dispersion relation and we investigate the equilibration of the
instabilities in section 5 through numerical integrations of the resulting S3T dynamical system. The predictions of
S3T are then compared to the results of the non-linear simulations in section 6 and we finally end with a brief
discussion of the obtained results and our conclusions in section 7.

\section{The emergence of coherent structures in non-linear simulations of a barotropic flow}

Consider a nondivergent barotropic flow on a $\beta$-plane with cartesian coordinates
$\boldsymbol{x}=(x,y)$. The velocity field, $\boldsymbol{u}=(u, v)$, is  given by
$(u,v)=(-\partial_y\psi,\partial_x\psi)$, where $\psi$ is the  streamfunction.
Relative vorticity $\zeta(x,y,t)= \Delta \psi$, evolves according to the non-linear (NL) equation:
\begin{equation}
\left ( \partial_t + \boldsymbol{u}\cdot\nabla \right )\zeta +\beta v=-r\zeta-\nu\Delta^2\zeta+f^e,\label{eq:derivation1}
\end{equation}
where $\Delta=\partial_{xx}^2+\partial_{yy}^2$ is the horizontal Laplacian, $\beta$ is the gradient of planetary vorticity,
$r$ is the coefficient
of linear dissipation that typically parameterizes  Ekman drag and $\nu$ is the coefficient of hyper-diffusion
that dissipates the energy flowing into unresolved scales. The forcing term
$f^e$ is necessary to sustain turbulence and  serves as a parameterization of  processes that are missing from the
barotropic dynamics, such as small scale
convection or baroclinic instability.  We will consider the flow to be on a doubly periodic channel
of size $2 \upi \times 2 \upi$.

As in many previous studies  the exogenous excitation $f^e$ will be
assumed to be a temporally delta correlated and spatially homogeneous
and isotropic random stirring
with a two-point, two-time correlation function of the form:
\begin{equation}
\left<f^e(x_1,y_1,t_1)f^e(x_2,y_2,t_2)\right>=\delta(t_2-t_1)\Xi(x_1, x_2, y_1, y_2),\label{eq:forc_prop}
\end{equation}
where the brackets denote an ensemble average over the different realizations of the forcing. The
temporally delta correlated stochastic forcing has the important property that
the  energy absorbed by the fluid is independent of the state of the flow and depends only on the
statistics of the forcing. The spatially homogeneous covariance of the forcing, $\Xi$, in the doubly
periodic channel can be written as the Fourier sum:
\begin{equation}
\Xi(x_1, x_2, y_1, y_2)=\sum_k \sum_l \hat{\Xi}(k, l)
\mathrm{e}^{\mathrm{i}k(x_1-x_2)+\mathrm{i}l(y_1-y_2)}, \label{eq:force_cov}
\end{equation}
with the  $x$, $y$ wavenumbers, $k$ and $l$, taking all integer values.  The Fourier amplitude
\begin{equation}
\hat{\Xi}(k, l)=\frac{\varepsilon K_f}{\Delta K_f}\left\{\begin{array}{ll} 1,~\mbox{for~}|\sqrt{k^2+l^2}-K_f|\leq \Delta
K_f\\0,~\mbox{for~}|\sqrt{k^2+l^2}-K_f|>\Delta K_f\end{array}\right. , \label{eq:finite_ring}
\end{equation}
is chosen so that the excitation injects energy at rate $\varepsilon$ in a narrow ring in wavenumber space with radius $K_f$
and width $\Delta K_f$.

Equation (\ref{eq:derivation1}) is solved using a pseudospectral code with a $128\times 128$ resolution and a fourth
order Runge-Kutta scheme for time stepping. While we vary the forcing energy input rate across a wide range of values,
the rest of the parameters are fixed at $\beta=10$, $r=0.01$, $\nu=1.19\cdot 10^{-6}$, $K_f=10$ and
$\Delta K_f=1$ yielding a non-dimensional beta parameter $\tilde\beta=\beta/(K_fr)=100$. As seen in Table 1,
showing the values of the non-dimensional beta parameters, $\tilde\beta$, and energy injection rates,
$\tilde\varepsilon=\varepsilon K_f^2/r^3$, for the Earth's atmosphere and ocean as well as for the Jovian
atmosphere, this choice of $\tilde\beta$ is relevant for both the Earth's ocean and the Jovian atmosphere.
\begin{table}
\footnotesize\caption{Relevant parameters in geophysical flows. Analysis for the values in the table is
given in Appendix A.}
\label{table1}\centering
\begin{tabular}{ p{3.5 cm} p{1.8 cm} p{1.8 cm} p{1.8 cm} p{1.3 cm} p{1.5 cm}}
 &  $1/K_f$ (km)   &   $1/r$ (days)  & $\varepsilon~(\mbox{m}^2\mbox{s}^{-3})$ & $\tilde\beta$ & $\tilde\varepsilon$\\
	\hline
Earth's atmosphere & 1000 & 10 & $3\cdot 10^{-4}$ & 15 & 190\\
\hline
Earth's ocean & 20 & 1000 & $10^{-9}$ & 40 & 2500\\
\hline
Jovian atmosphere & 100 & 5800 & $0.5\cdot 10^{-5}$ & 125 & $1.25\cdot 10^{11}$\\
\end{tabular}
\end{table}

\begin{figure}
\centerline{\includegraphics[width=6in]{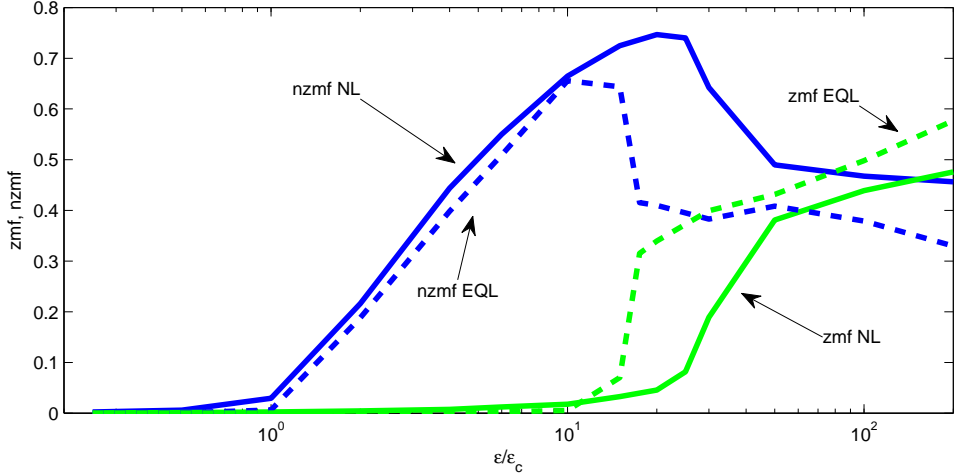}}
\caption{The \mbox{zmf} and \mbox{nzmf} indices defined in (\ref{eq:zmf}) and
(\ref{eq:nzmf}) respectively, as a function of energy input rate $\varepsilon/\varepsilon_c$ for the non-linear
(NL) integrations and the ensemble quasi-linear (EQL) integrations (dashed line) with $N_{ens}=10$ ensemble members
as described in section 6. The critical value $\varepsilon_c=8.4\cdot 10^{-6}$ is the energy input rate at which the S3T predicts
structural instability of the homogeneous turbulent state. Zonal jets emerge for $\varepsilon > \varepsilon_{nl}$, with
$\varepsilon_{nl} =15\varepsilon_c$. The parameters are $\beta=10$, $r=0.01$, $\nu=1.19\cdot 10^{-6}$ and the forcing
is an isotropic ring in wavenumber space with radius $K_f=10$ and width $\Delta K_f = 1$.}
\label{fig:zmf}
\end{figure}

\begin{figure}
\centerline{\includegraphics[width=6in]{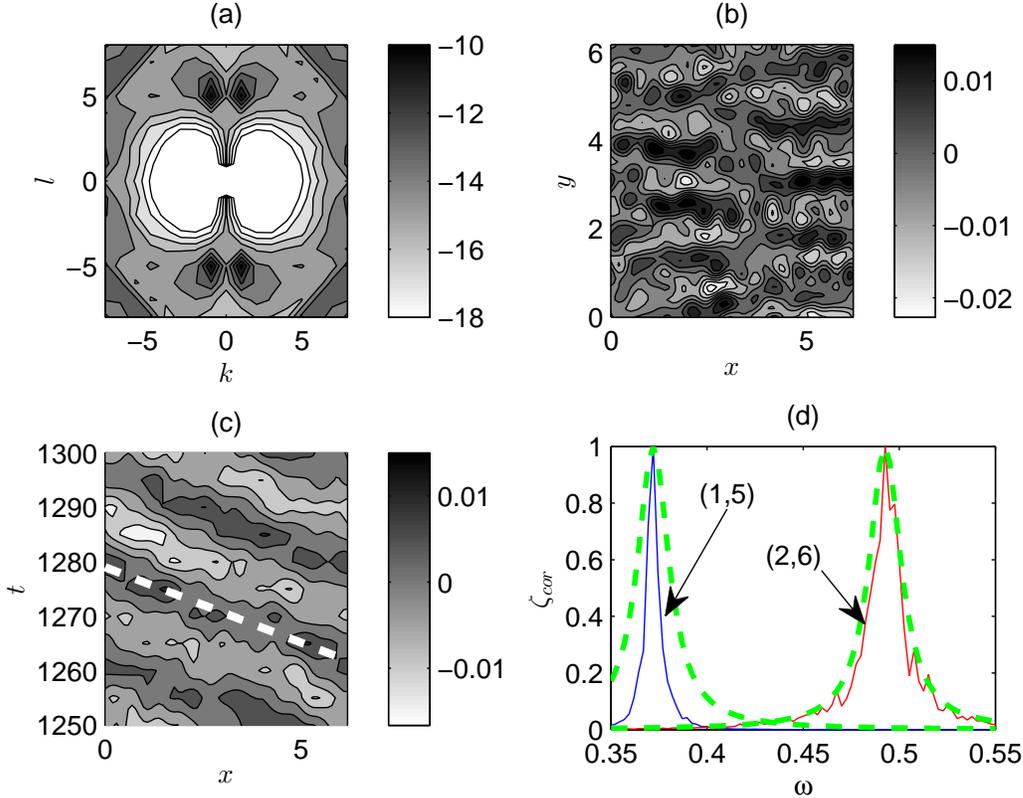}}
\vspace*{-5mm}
\caption{(a) Time averaged energy power spectra, $\log(\hat{E}(k,l))$, obtained from the non-linear
(NL) simulation of (\ref{eq:derivation1}) at $\varepsilon/ \varepsilon_c=4$. The flow is dominated by a
$(|k|,|l|)=(1,5)$ non-zonal coherent structure that is evident in the snapshot of the streamfunction
$\psi(x,y,t)$ (panel b) and the Hovm\"oller diagram of $\psi(x,y=\upi/4,t)$ (panel c). The thick dashed line
in (c) corresponds to the phase speed obtained from the eigenvalue relation (\ref{eq:dispersion}). (d) The
ensemble mean wavenumber--frequency power spectrum $\zeta_{cor}(\omega, k, l)$ as a function of frequency
for $(k, l)=(1, 5)$ and $(k, l)=(2, 6)$. The corresponding spectrum $\zeta_{cor}^R$ for stochastically
forced linear Rossby waves that remain phase coherent over $1/r$ is also shown (dashed lines). All correlation
functions are normalized to one to facilitate comparison.}
\label{fig:NL_snap1}
\end{figure}

The non-linear
system reaches a statistical equilibrium at about $t=10/r$. Following previous studies \citep{Galperin-etal-06}, the integration
was carried until $t=100/r$ in order to collect accurate statistics and the last $80/r$ time units were used for calculating the time
averages. To illustrate some of the characteristics of the
turbulent flow and the emergence of structure, we consider two indices that measure the power which is concentrated at scales larger than the scales forced.
The first is the zonal mean flow index defined as   in
\cite{Srinivasan-Young-12}, as  the ratio of the energy of zonal jets with scales larger than the scale of the forcing over the total energy
\begin{equation}
\mbox{zmf}=\frac{\sum_{l:l<K_f-\Delta K_f} \hat{E}(k=0,l)}{\sum_{kl}\hat{E}(k,l)},\label{eq:zmf}
\end{equation}
where $\hat{E}(k,l)$ is the time averaged energy power spectrum of the flow at wavenumbers $(k,l)$. The second is the non-zonal mean flow index
defined as the ratio of the energy of the non-zonal modes with scales larger than the scale of the forcing over the total
energy:
\begin{equation}
\mbox{nzmf}=\frac{\sum_{kl: K<K_f-\Delta K_f}\hat{E}(k,l)}
{\sum_{kl}\hat{E}(k,l)}-\mbox{zmf}.\label{eq:nzmf}
\end{equation}
If the structures that emerge are coherent, then these indices quantify their amplitude. Figure \ref{fig:zmf} shows both indices as a
function of the energy input rate $\varepsilon$. Remarkably, both indices exhibit sharp increases at critical energy input rates, indicating
the occurrence of regime transitions  in the flow.
For $\varepsilon$ smaller than the critical value $\varepsilon_c$, the turbulent
flow is homogeneous and remains translationally
invariant in both directions and both indices are nearly zero. When $\varepsilon>\varepsilon_c$, non-zonal
structures that have scales larger than the scale of the forcing form, as indicated by the increase in the
$\mbox{nzmf}$ index. The critical value is estimated from the point of rapid increase of the nzmf index to be
$\varepsilon_c=8.4\cdot 10^{-6}$ (for the parameters chosen) but this value is also verified by the S3T stability
analysis in section 4. The time averaged power spectrum shown in figure \ref{fig:NL_snap1}(a) for
$\varepsilon=4\varepsilon_c$, is anisotropic with a pronounced peak at $(|k|, |l|)=(1, 5)$. This peak corresponds
to a structure with the corresponding scale that is evident in the vorticity field evolution. This is illustrated by
the appearance of a structure with $(|k|, |l|)=(1, 5)$ in the snapshot of the streamfunction field shown in figure
\ref{fig:NL_snap1}(b). The Hovm\"oller diagram in which contours of $\psi(x,y=\upi/4,t)$ are plotted in figure
\ref{fig:NL_snap1}(c) shows that this structure is coherent and propagates in the retrograde direction. The sloping
dashed line in the diagram corresponds to the phase speed of the waves, which is found to be approximately the Rossby
wave phase speed for $(k, l)=(1, 5)$. We obtain an estimate of the phase coherence of this structure by calculating
the ensemble mean of the wavenumber--frequency power spectrum of its vorticity field:
\begin{equation}
\zeta_{cor}(\omega, k, l)=\left<\left|\hat\zeta (k, l, \omega)\right|^2\right>,\label{eq:Ucor}
\end{equation}
where
\begin{equation}
\hat\zeta(k, l, \omega)=\int\sum_{x_i}\sum_{y_i}\zeta(x_i, y_i, t)\mathrm{e}^{-\mathrm{i}kx_i-\mathrm{i}ly_i-i\omega t}dt.
\end{equation}
Traveling wave structures manifest as peaks of $\zeta_{cor}$ at specific frequencies with a half-width proportional to the
time scale of their phase coherence. For example, for linear Rossby waves that are stochastically forced and damped with
rate $1/r$:
\begin{equation}
\zeta_{cor}^R(\omega, k, l)\sim \frac{1}{\left[\omega-\beta k/(k^2+l^2)\right]^2+r^2},\label{eq:Ucor_ross}
\end{equation}
and the waves are phase correlated over the dissipation time scale \citep{Galperin-etal-10}. We will consider the
structures in the nonlinear simulation to be phase coherent when their coherence time exceeds $1/r$. Figure
\ref{fig:NL_snap1}(d) shows the ensemble mean power spectrum $\zeta_{cor}(\omega, k, l)$ as obtained from
the nonlinear simulations for two structures, along with the corresponding power spectrum $\zeta_{cor}^R$
of half-width $1/r$ for the same waves. The dominant $(|k|, |l|)=(1, 5)$ structure
is coherent over about four dissipation time scales, whereas the other less prominent structures (as for
example the $(|k|, |l|)=(2, 6)$ shown), are coherent over the dissipation time scale, as if stochastically forced.
The $(|k|, |l|)=(1, 5)$ structure dominates the flow (with 60\% of the total energy concentrated in this
structure) and remains coherent up to $\varepsilon/ \varepsilon_c<15$. Therefore  the increase in the $\mbox{nzmf}$ index
observed in figure \ref{fig:zmf} signifies the emergence of non-zonal coherent structures that break the
translational symmetry of the turbulent state simultaneously in both the $x$ and $y$ direction.

\begin{figure}
\centerline{\includegraphics[width=6in]{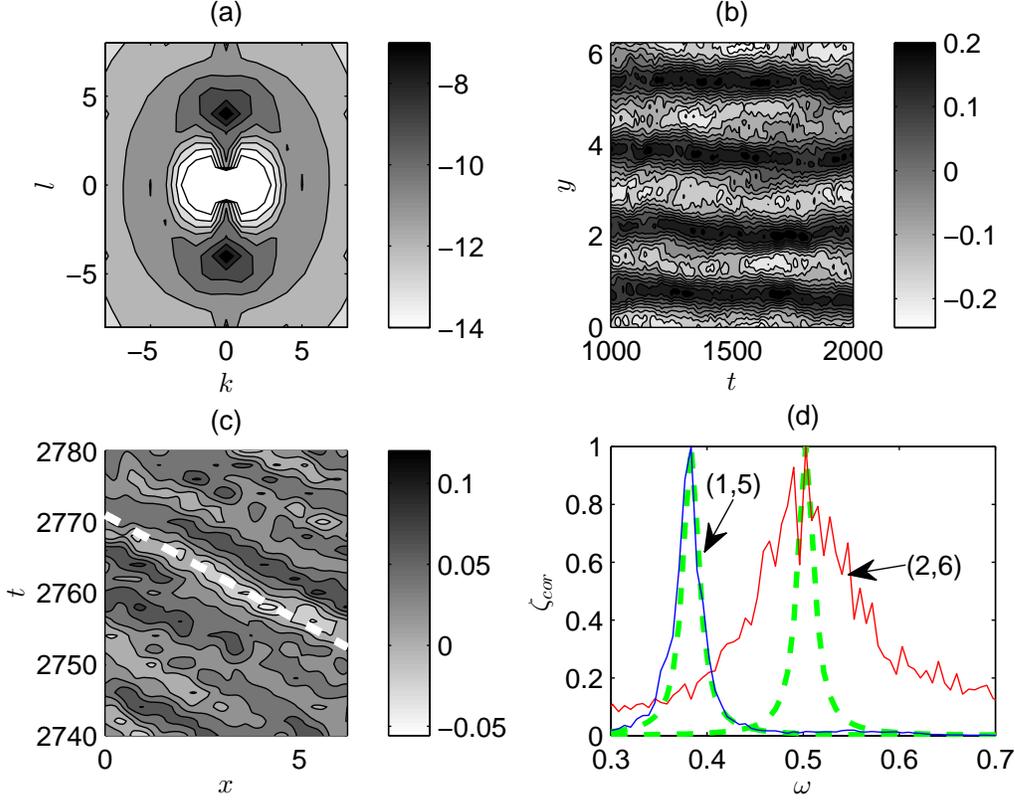}}
\vspace*{-5mm}
\caption{(a) Time averaged energy power spectra, $\log(\hat{E}(k,l))$, obtained from the non-linear
simulation of (\ref{eq:derivation1}) at $\varepsilon/ \varepsilon_c=50$. The flow is dominated by a
$(k,|l|)=(0,4)$ zonal jet that is evident in the Hovm\"oller diagram of the $x$-averaged $\overline{\psi(x,y,t)}$
(panel b). (c) Hovm\"oller diagram of $\psi(x,y=\upi/4,t)$ showing the presence of westward propagating structures.
The thick dashed line corresponds to the phase speed obtained from (\ref{eq:dispersion}). (d) The
ensemble mean wavenumber--frequency power spectrum $\zeta_{cor}(\omega, k, l)$ as a function of frequency
for $(k, l)=(1, 5)$ and $(k, l)=(2, 6)$. The corresponding spectrum $\zeta_{cor}^R$ for stochastically
forced linear Rossby waves is also shown (dashed lines). All correlation functions are normalized to one to
facilitate comparison.}
\label{fig:NL_snap2}
\end{figure}

The rapid increase in the $\mbox{zmf}$ index shown in  figure \ref{fig:zmf} above
$\varepsilon=15\varepsilon_c\equiv\varepsilon_{nl}$, indicates a second regime transition in the flow with the
emergence of robust and coherent zonal jets.  For $\varepsilon=3.3\varepsilon_{nl}$  (i.e. $\varepsilon/ \varepsilon_c=50$)
the spectrum,  shown in  figure \ref{fig:NL_snap2}(a), has  significant power at the zonal structures  with $(k,|l|)=(0, 4)$.
These peaks correspond to coherent zonal jets as illustrated by the Hovm\"oller diagram of the zonally averaged
streamfunction shown in figure \ref{fig:NL_snap2}(b). However, there is significant power in non-zonal
structures (in this case with wavenumbers $(|k|,|l|)=(1, 4)$ and $(|k|,|l|)=(1, 5)$), a characteristic that is
also revealed by the high values of the nzmf index for large energy input rates. The Hovm\"oller diagram and
the ensemble mean power spectrum $\zeta_{cor}(\omega, k, l)$ shown in figures \ref{fig:NL_snap2}(c) and
\ref{fig:NL_snap2}(d), reveal that the non-zonal structures are propagating in the retrograde direction and remain
coherent over at least a dissipation time scale, whereas the peaks of $\zeta_{cor}(\omega, k, l)$ at other
structures have been significantly broadened by turbulence. The phase speed calculated from the diagram is
different from the corresponding Rossby wave speed for both $(|k|,|l|)=(1, 4)$ and $(|k|,|l|)=(1, 5)$. At larger energy input
rates the zonal jets have typically larger scales due to jet merging and coexist with energetically significant
westward propagating non-zonal structures having an energy
between $10-50 \%$ of the jet energy and scales $(|k|, |l|)=(1, m)$, where $m$ is the number of jets in the channel.
However the phase coherence of these waves is
a decreasing function of $\varepsilon$.

Similar Rossby-like, westward
propagating coherent structures were also reported recently in numerical simulations of the barotropic vorticity equation on the sphere
\citep{Sukoriansky-etal-2008,Galperin-etal-10}. In agreement with the results presented in this
work these large scale waves contain a significant amount of energy. In the regime in which zonal jets are absent or weak
these waves were found to follow the Rossby wave dispersion. In the regime
in which strong zonal jets dominate the flow (called the zonostrophic regime by these authors), the waves propagate with markedly
different phase speeds. These waves were therefore
classified as linear Rossby waves in the former and as satellite modes \citep{Danilov-04} or zonons
\citep{Sukoriansky-etal-2008} in the latter regime.

We will show next that the emergence and characteristics of both the zonal and the non-zonal coherent structures
can be accurately predicted by considering the stability of a particular second order closure of the turbulent dynamics.
This second order closure results in a non-equilibrium statistical theory,  called Stochastic Structural
Stability Theory (S3T) or Second Order Cumulant Expansion theory (CE2)
\citep{FI-03,FI-07,Marston-etal-2008,Bakas-Ioannou-2011,Srinivasan-Young-12, Marston-2012}, that addresses
the emergence of structure in planetary turbulence.

\section{Formulation of Stochastic Structural Stability Theory}

S3T describes the statistical dynamics of the first two equal time cumulants of (\ref{eq:derivation1}). The first
cumulant is the ensemble mean of the vorticity $Z(\mathbf{x},t) \equiv\left< \zeta\right>$. The second cumulant
$C(\mathbf{x}_1, \mathbf{x}_2, t)\equiv \left<\zeta_1'\zeta_2'\right>$, is the two point correlation function of the
vorticity deviation from the mean $\zeta_i'\equiv\zeta_i-Z_i$. We use the
shorthand $\zeta_i=\zeta(\mathbf{x}_i, t)$, with $i=1,~2$ to refer to the value of the relative vorticity at the
specific point $\mathbf{x}_i=(x_i, y_i)$. In most earlier studies of S3T,  the ensemble average was assumed to represent a zonal
average. With this interpretation of the ensemble average,  the  non-zonal structures
are  treated as incoherent  motions
and  the theory  can only address the emergence of  zonal jets.
In order to address the emergence of coherent non-zonal structures in turbulence,
we adopt in this work the more general interpretation that the ensemble average is a Reynolds average
over the fast turbulent motions that typically have time scales in this case $\tau_{eddy}\ll 1/r$. The averaging
time scale is taken to be several eddy decorrelation scales but also smaller than $1/r$ and smaller
than the period of the propagating structures (for a periodic box the lowest period is of order $1/ \beta$) in order
to retain the slow evolution of the coherent structures. This interpretation of the S3T has been adopted recently in
studies of non-zonal blocking patterns in baroclinic two-layer turbulence by \cite{Bernstein-2009} and
\cite{Bernstein-Farrell-2010}. With this definition of the ensemble mean, we seek to obtain the statistical dynamics
of the interaction of the coarse-grained ensemble average field, which can be zonal or  non-zonal coherent structures
represented by their vorticity $Z$, with the fine-grained incoherent field  represented by the vorticity second
cumulant $C$.

The equations governing the evolution of the first two cumulants are obtained as follows. Under the decomposition
of vorticity into an ensemble mean and a deviation from the mean, (\ref{eq:derivation1}) is split into two equations governing
the evolution of the eddy (deviation from the mean) vorticity $\zeta'$ and the vorticity of the coherent structures $Z$:
\begin{equation}
\left(\partial_t+\boldsymbol{U}\cdot\nabla\right)\zeta' +(\beta+\partial_yZ)v'+u'\partial_xZ=-r \zeta'
-\nu\Delta^2\zeta'+\underbrace{f^e + f^{nl}}_{f} ,\label{eq:q_evo}
\end{equation}
\begin{equation}
\left(\partial_t+\boldsymbol{U}\cdot\nabla\right)Z+\beta V=-\nabla\cdot\left<\boldsymbol{u}'\zeta'\right>-rZ-
\nu\Delta^2Z,\label{eq:Q_evo}
\end{equation}
where $\boldsymbol{u}'=[u', v']=[-\partial_y\psi ', \partial_x\psi ']$ and $\boldsymbol{U}=[U, V]=
[-\partial_y\Psi, \partial_x\Psi]$ are the non-divergent eddy and ensemble mean velocity fields,
\begin{equation}
f^{nl}=\left< \boldsymbol{u}'\cdot\nabla \zeta'\right>- \boldsymbol{u}'\cdot\nabla \zeta',\label{eq:eddy-eddy}
\end{equation}
is the forcing term from the non-linear interactions among the turbulent eddies and $f=f^e+f^{nl}$ represents
the total eddy forcing. The ensemble average vorticity fluxes $\left< \mathbf{u}'\zeta'\right>$ can be expressed in terms of the
second cumulant of vorticity as:
\begin{equation}
\left< \mathbf{u}'\zeta'\right>=\left[\left<u_1'\zeta_2'\right>_{\mathbf{x}_1=\mathbf{x}_2},
\left<v_1'\zeta_2'\right>_{\mathbf{x}_1=\mathbf{x}_2}\right]=\left[-\left(\partial_{y_1}\Delta_1^{-1}
C\right)_{\mathbf{x}_1=\mathbf{x}_2},\left(\partial_{x_1}\Delta_1^{-1}
C\right)_{\mathbf{x}_1=\mathbf{x}_2}\right],\label{eq:vor_fluxes1}
\end{equation}
where $\Delta^{-1}$ is the integral operator that inverts vorticity into the streamfunction field
($\psi=\Delta^{-1}\zeta$). The subscripts in the operators in (\ref{eq:vor_fluxes1}) denote the variable
$\mathbf{x}_i$ on which the operators act. For example $\partial_{x_i}$ denotes differentiation
with respect the variable $x_i$ ($i=1,2$), while the integral operators $\Delta_i^{-1}$
invert the vorticity covariance with respect to variables $\mathbf{x}_i$ so that the streamfunction
covariance is $S(\mathbf{x}_1, \mathbf{x}_2)=\Delta^{-1}_1 \Delta^{-1}_2 C$. The subscript
$\mathbf{x}_1 = \mathbf{x}_2$ means that the expression in parenthesis is calculated at the same point.
As a result, the first cumulant evolves as:
\begin{equation}
\partial_t Z+UZ_x+V(\beta+Z_y)=\partial_x\left(\partial_{y_1}\Delta_1^{-1}
C\right)_{\mathbf{x}_1=\mathbf{x}_2}-\partial_y\left(\partial_{x_1}\Delta_1^{-1}
C\right)_{\mathbf{x}_1=\mathbf{x}_2}-rZ-\nu\Delta^2Z.\label{eq:Q_evo2}
\end{equation}
Multiplying (\ref{eq:q_evo}) for
$\partial_t\zeta_1'$ by $\zeta_2'$ and (\ref{eq:q_evo}) for $\partial_t\zeta_2'$ by $\zeta_1'$, adding the two
equations and taking the ensemble average yields:
\begin{equation}
\partial_t C=(A_1+A_2)C+\left<f_1\zeta_2'+f_2\zeta_1'\right>,\label{eq:cov_evo1}
\end{equation}
where
\begin{equation}
A_i=-U_i\partial_{x_i}-V_i\partial_{y_i}-(\beta+\partial_{y_i}Z)\partial_{x_i}\Delta_i^{-1}+\partial_{x_i}Z
\partial_{y_i}\Delta_i^{-1}-r-\nu\Delta_i^2,\label{eq:op_A}
\end{equation}
governs the dynamics of linear perturbations about the instantaneous mean flow $\boldsymbol{U}$,
\begin{eqnarray}
\left<f_1\zeta_2'+f_2\zeta_1'\right>&=&\left<f_1^e\zeta_2'+f_2^e\zeta_1'\right>+
\left<f_1^{nl}\zeta_2'+f_2^{nl}\zeta_1'\right>\nonumber\\ &=&\left<f_1^e\zeta_2'+f_2^e\zeta_1'\right> +\nonumber \\
&+&
\left[\left(\partial_{y_1x_3}^2-\partial_{x_1y_3}^2\right)\Delta_2^{-1}\Gamma\right]_
{\mathbf{x}_1=\mathbf{x}_3} + \left[\left(\partial_{y_2x_3}^2-\partial_{x_2y_3}^2\right)
\Delta_2^{-1}\Gamma\right]_{\mathbf{x}_2=\mathbf{x}_3},\label{eq:third_cum}
\end{eqnarray}
and $\Gamma\equiv \left<\zeta_1'\zeta_2'\zeta_3'\right>$ is the third cumulant. The first term on the right
hand side of (\ref{eq:third_cum}) is the correlation of the external forcing $f^e$ with vorticity,
while the other two terms involve the third cumulant that describes the eddy-eddy interactions. Previous
studies addressing the interaction of turbulent eddies with zonal jets in baroclinic turbulence, as
well as the interaction of coherent vortices with small scale turbulence, have shown that several important
features of the coherent flow as well as accurate eddy statistics are obtained by either neglecting
or suitably parameterizing the eddy-eddy non-linearity $f^{nl}$ as stochastic forcing and enhanced dissipation
\citep{FI-93d,Sole_Farrell-96,Dubrulle-Nazarenko-1997,Laval-etal-2000,DelSole-04,OGorman-Schneider-2007,Marston-etal-2008}. This
is equivalent to setting the
third cumulant to zero, or parameterizing the last two terms in (\ref{eq:third_cum}) as a given
correlation function.
We note that the distinction between these two parameterizations is semantic for barotropic turbulence sustained by
stochastic forcing. However, if turbulence is self-maintained without any external stochastic forcing, as for example is the case in
baroclinic flows, and the third cumulant is altogether neglected then the covariance in (\ref{eq:cov_evo1}) is
unforced and will evolve to the low rank structure of the Lyapunov vector of the generally time dependent $A$
operator. As a result, it will fail to accurately represent
the second order statistics of the turbulent flow \citep{Marston-etal-2008}. The presence of the parameterization
of the nonlinear eddy-eddy scattering as noise in this case, is therefore important because it keeps the structure of
the second order cumulant full rank and in accordance to the amplification properties of the non-normal $A$ operator.
In this work we will neglect the third-order cumulants and show that the S3T theory with this approximation can
accurately predict the emergence of large scale structures in the flow.
We therefore assume  that $f$ is the delta correlated
external forcing, $f^e$, with  $\left<f_1\zeta_2'+f_2\zeta_1'\right>=\left<f_1f_2\right>=\Xi$. With this approximation
the second order statistics evolve according to:
\begin{equation}
\partial_t C=(A_1+A_2)C+\Xi.\label{eq:cov_evo2}
\end{equation}

Equations (\ref{eq:Q_evo2}) and (\ref{eq:cov_evo2}) form a closed deterministic system that governs the
joint evolution of the coherent flow field and of the second order turbulent eddy statistics. This second
order closure is the basis of Stochastic Structural Stability Theory \citep{FI-03}.
The S3T system can have fixed points, limit cycles or chaotic attractors.
Examples of the attractor of this system can be found in the S3T description of the organization of
geophysical and plasma turbulence into zonal jets \citep{FI-03,FI-08,Farrell-Ioannou-2009-plasmas}, as
well as in the S3T description of blocking patterns in the atmosphere \citep{Bernstein-Farrell-2010}.
The fixed points $Z^E$ and $C^E$, if they exist, define statistical equilibria of the coherent structures
with vorticity, $Z^E$, in the presence of an eddy field with covariance, $C^E$. The structural stability
of these turbulent equilibria that can be investigated in S3T, addresses the parameters in the physical
system which can lead to abrupt reorganization of the turbulent flow. Specifically, when an equilibrium of
the S3T equations becomes unstable as a physical parameter changes, the turbulent flow bifurcates to a
different attractor. In this work, we show that coherent structures emerge as unstable modes of the S3T system and
equilibrate at finite amplitude. The predictions of the S3T system regarding the emergence and characteristics
of the coherent structures are then compared to the non-linear simulations.

\section{S3T instability and emergence of finite amplitude large scale structure}

The homogeneous equilibrium with no mean flow
\begin{equation}
Z^E=0,~~C^E={\Xi\over 2r},\label{eq:equil}
\end{equation}
is a fixed point of the S3T system  (\ref{eq:Q_evo2}) and  (\ref{eq:cov_evo2}) in the absence of hyperdiffusion
(cf. Appendix B). The stability of this
homogeneous equilibrium, can be addressed by performing eigenanalysis of the S3T system linearized about the equilibrium.
Because of the absence of coherent mean flow and the homogeneity of
$C^E$ we can seek  eigensolutions in the modal  form
$\delta Z=Z_{nm}\mathrm{e}^{\mathrm{i}nx+\mathrm{i}my}\mathrm{e}^{\sigma t}$ and $\delta C = C_{nm}(\tilde{x}, \tilde{y})
\mathrm{e}^{\mathrm{i}n\overline{x}+\mathrm{i}m\overline{y}}\mathrm{e}^{\sigma t}$, where $\tilde{x}=x_1-x_2$,
$\overline{x}=(x_1+x_2)/2$,
$\tilde{y}=y_1-y_2$, $\overline{y}=(y_1+y_2)/2$, $n$ and $m$ are the $x$ and $y$
wavenumbers of the eigenfunction and $\sigma=\sigma_r+\mathrm{i}\sigma_i$ is the eigenvalue with
$\sigma_r=\Real (\sigma)$, $\sigma_i=\Imag (\sigma)$ being the growth rate and frequency of the
mode respectively. The eigenvalue $\sigma$  satisfies the equation:
\begin{eqnarray}
& &\int_{-\infty}^{\infty}\int_{-\infty}^{\infty}\frac{(mk-nl)\left[nm(k_+^2-l_+^2)+(m^2-n^2)k_+l_+
\right](1-N^2/K^2)\hat{\Xi}(k,l)}{2\mathrm{i}\beta k_+(k_+n+l_+m)-\mathrm{i}n\beta\left(K^2+K_s^2\right)/2+(\sigma+2r)K^2
K_s^2}\mathrm{d}k\mathrm{d}l\nonumber\\ &=&2r\upi(\sigma+r)N^2-2r\mathrm{i}\upi n\beta,\label{eq:dispersion}
\end{eqnarray}
where
\begin{equation}
\hat{\Xi}(k,l)=\frac{1}{2\upi} \int_{-\infty}^{\infty}\int_{-\infty}^{\infty}
\Xi(\tilde{x},\tilde{y})\mathrm{e}^{-\mathrm{i}k\tilde{x}-\mathrm{i}l\tilde{y}}
\mathrm{d}\tilde{x}\mathrm{d}\tilde{y}~,\label{eq:eq_stream_cov}
\end{equation}
is the Fourier transform of the forcing covariance, $K^2=k^2+l^2$,
$K_s^2=(k+n)^2+(l+m)^2$, $N^2=n^2+m^2$, $k_+=k+n/2$ and $l_+=l+m/2$ (cf. Appendix B). For  zonally
homogeneous perturbations with  $n=0$, (\ref{eq:dispersion}) reduces to
the eigenvalue relation derived by \cite{Srinivasan-Young-12} for the emergence of jets in a
barotropic $\beta$-plane. Eigenvalue  relation (\ref{eq:dispersion}) was derived for a  flow  that extends to infinity.
For the periodic channel considered in the non-linear simulations, the corresponding  eigenvalue relation
is readily obtained by substituting the integrals in (\ref{eq:dispersion}) and (\ref{eq:eq_stream_cov}) with summation
over integer values of $k$ and $l$ \citep{Bakas-Ioannou-2013}. We non-dimensionalize the eigenvalue relation using the dissipation time
scale $1/r$ and a typical forcing length scale $L_f$ and rewrite (\ref{eq:dispersion}) in the general form:
\begin{equation}
\tilde{\sigma}_{(\tilde{n}, \tilde{m})} = g( \tilde{\beta}, \tilde{\varepsilon}).\label{eq:nondim_disp}
\end{equation}
For a given spectral distribution of the forcing, (\ref{eq:nondim_disp}) gives the eigenvalue $\tilde{\sigma}=\sigma/r$ for each
wavenumbers $(\tilde n,  \tilde m)= L_f (n,m)$ as a function of the planetary vorticity gradient $\tilde{\beta}=\beta L_f /r$
and the energy injection rate $\tilde{\varepsilon}=\varepsilon / ( r^3 L_f^2)$.

\begin{figure}
\centerline{\includegraphics[width=5in]{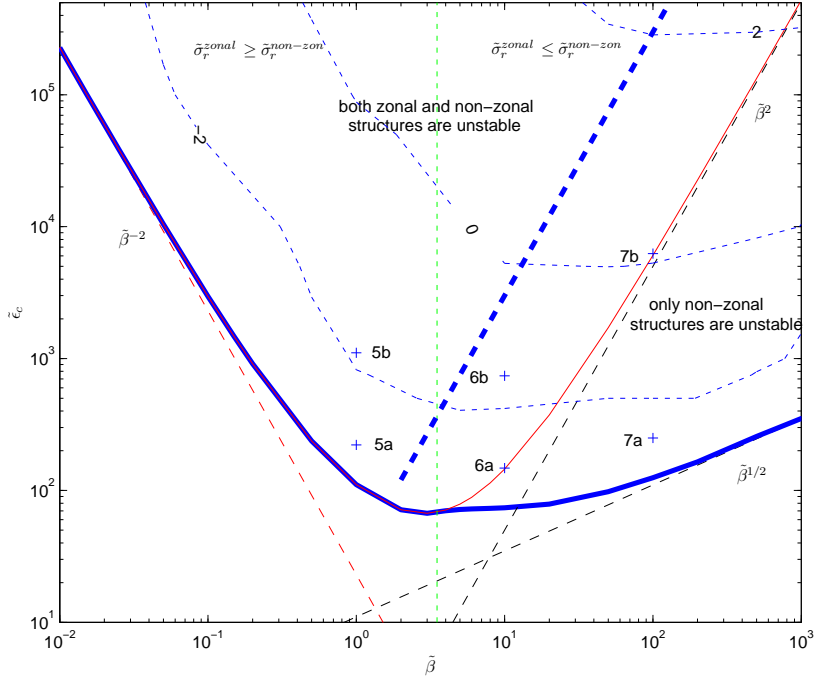}}
\caption{The critical energy input rate $\tilde \varepsilon_c$ for structural instability (thick solid line) and
the critical energy input rate for structural instability of zonal jets (solid line) as a function of
$\tilde{\beta}$. The behavior of these critical values for large and small $\tilde \beta$ is  indicated with the dashed asymptotes
$\tilde \varepsilon = 23 \tilde \beta^{-2}$ for $\tilde \beta \ll 1$,   $\tilde \varepsilon = 11 \tilde \beta^{1/2}$ and
$\tilde \varepsilon = 0.5 \tilde \beta^{2}$  for the emergence of non-zonal and zonal structures respectively for
$\tilde \beta\gg 1$.  Above the thick (thin) line non-zonal (zonal) coherent structures emerge. The thin dotted vertical line
$\tilde \beta =\tilde \beta_{min}$ separates the unstable region: for $\tilde \beta<\tilde \beta_{min}$ the zonal structures grow
the most, whereas for $\tilde \beta> \tilde \beta_{min}$ the non-zonal structures grow the most.  Also shown are the contours (thick
dashed lines) of the  maximum  growth rate $\tilde\sigma_{max}$ (with contour values corresponding to $\log(\tilde\sigma_{max})$).
The thick dotted line $\tilde\varepsilon = 30 \tilde \beta^2$ is the locus of the points on which the maximum $\tilde \sigma_r$ occurs
for each $\tilde \varepsilon$. The  crosses indicate  the  $\tilde \varepsilon$ and $\tilde \beta$ values for which the dispersion relation
of the unstable modes is shown in  figures 5-7.}
\label{fig:emin}
\end{figure}

We consider the case of a ring forcing that injects energy at rate $\varepsilon$ at the total wavenumber $K_f$:
\begin{equation}
\hat{\Xi}(k,l)= 2 \varepsilon K_f \delta(\sqrt{k^2+l^2}-K_f),
\end{equation}
which is an idealization of the forcing (\ref{eq:finite_ring}) used in the non-linear simulations. We then obtain
the eigenvalues $\tilde\sigma$ for an infinite domain by numerically solving (\ref{eq:nondim_disp}).
For small values of the energy
input rate, the growth rate $\tilde{\sigma}_r$ is negative for all $(\tilde n,\tilde m)$ and the homogeneous equilibrium
is stable. At a critical $\tilde \varepsilon_c$ the homogeneous flow becomes S3T unstable, symmetry breaking occurs
and exponentially growing coherent structures  emerge. The critical value, $\tilde \varepsilon_c$,  is calculated by first
determining the energy input rate $\tilde{\varepsilon}_t(\tilde{n}, \tilde{m})$
that renders wavenumbers $(\tilde{n}, \tilde{m})$
neutral $\left(\tilde{\sigma}_{r (\tilde{n}, \tilde{m})} =0\right)$, and then by  finding the minimum  energy input rate
over all wavenumbers: $\tilde \varepsilon_c=\mbox{min}_{(\tilde{n}, \tilde{m})} \tilde{\varepsilon}_t$. The critical
energy input rate $\tilde \varepsilon_c$ as a function of $\tilde{\beta}$ is shown in figure  \ref{fig:emin}. The absolute
minimum energy input rate required is
$\tilde \varepsilon_c=67$ and occurs at $\tilde{\beta}_{min}=3.5$. For $\tilde{\beta}\leq \tilde{\beta}_{min}$,
the structures that first become marginally stable  are zonal jets (with  $n=0$). The critical input
rate increases as $\tilde \varepsilon_c\sim \tilde{\beta}^{-2}$ for $\tilde{\beta} \rightarrow 0 $ (in agreement with the
findings of \cite{Srinivasan-Young-12}) and
the homogeneous equilibrium is structurally stable for all excitation amplitudes when $\tilde \beta =0$.
The structural stability for $\tilde \beta=0$ is an artifact of the assumed isotropy of the excitation and
in the presence of anisotropy the critical input rate, $\tilde \varepsilon_c$,  saturates
to a finite value as $\tilde\beta\rightarrow 0$ \citep{Bakas-Ioannou-2011,Bakas-Ioannou-2012}.
%
For
$\tilde{\beta}>\tilde{\beta}_{min}$, the marginally stable structures are non-zonal and $\tilde \varepsilon_c$ grows as
$\tilde \varepsilon_c\sim \tilde{\beta}^{1/2}$ for $\tilde{\beta} \rightarrow  \infty$. Since the critical forcing for the
emergence of zonal jets (also shown in figure \ref{fig:emin}), increases as $\tilde \varepsilon_c\sim \tilde{\beta}^2$ for
$\tilde{\beta} \rightarrow  \infty$
\citep{Srinivasan-Young-12}, for large values of $\tilde{\beta}$ non-zonal structures first emerge and only at
significantly higher $\tilde \varepsilon$ zonal jets are expected to appear. Investigation of these results with other
forcing distributions revealed that these results are independent of the isotropy of the forcing. Contours of the maximum growth rate of the
S3T instability, $\tilde\sigma_{max}=\mbox{max}_{(\tilde{n}, \tilde{m})} \tilde\sigma_r$ are also shown in
figure \ref{fig:emin} as a function of $(\tilde\varepsilon, \tilde\beta)$. For a given $\tilde\beta$, the maximum growth
rate increases monotonically with larger energy input rates, while for a given level of excitation $\tilde\varepsilon_m$
the  maximum growth rate occurs for a finite $\tilde\beta_m$ that satisfies roughly $\tilde\varepsilon_m\sim 30
\tilde\beta_m^2$ (represented by the thick dotted line in the figure).

For $\tilde{\varepsilon}>\tilde \varepsilon_c$  there is a number of structures that grows exponentially.
It is shown in Appendix B that for the isotropic forcing considered and for $n\neq 0$, the eigenvalues satisfy
the relations:
\begin{equation}
\tilde{\sigma}_{(-\tilde{n}, \tilde{m})}=\tilde{\sigma}_{(\tilde{n},
\tilde{m})}^*, \mbox{~and~}\tilde{\sigma}_{(\tilde{n}, -\tilde{m})}=\tilde{\sigma}_{(\tilde{n}, \tilde{m})}, \label{eq:symmet}
\end{equation}
implying  that the
growth rates depend on  $|\tilde{n}|$ and $|\tilde{m}|$.
As a result, the plane wave $\delta Z=\cos(nx+my)$ and an array of localized vortices $\delta Z=\cos (nx)\cos(my)$, have
the same growth rate, despite their different structure.

\begin{figure}
\centerline{\includegraphics[width=5in]{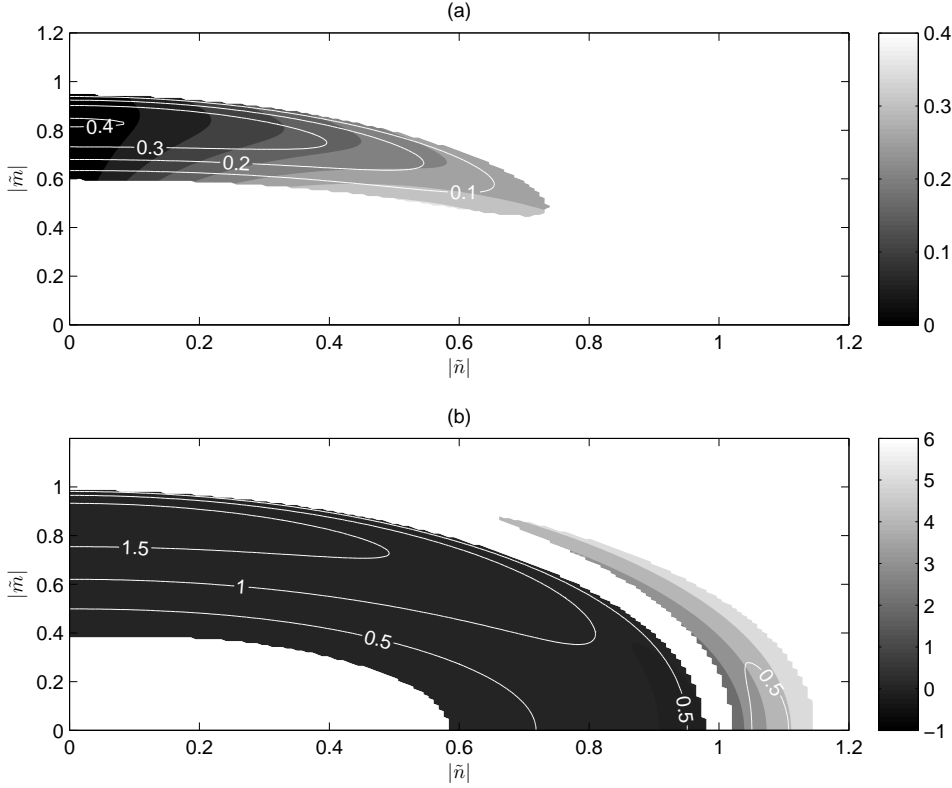}}
\caption{Dispersion relation of the unstable modes for $\tilde{\beta}=1$. The energy input rate is
(a) $\tilde{\varepsilon}=2\tilde \varepsilon_c$ and (b) $\tilde{\varepsilon}=10\tilde \varepsilon_c$. The contours
show the growth rate $\tilde{\sigma}_r$ and the shading shows the frequency $\tilde{\sigma}_i$ of the unstable modes.}
\label{fig:growth_b1_l0}
\end{figure}

We first consider the case  $\tilde{\beta}=1$,  $\tilde \varepsilon=2\tilde \varepsilon_c$, corresponding to
the point marked as \ref{fig:growth_b1_l0}a  in the $(\tilde \varepsilon, \tilde \beta)
$ regime diagram shown in figure \ref{fig:emin}. The growth rate of the maximally growing eigenvalue,
$\tilde{\sigma}_r$,  and its associated   frequency of
the mode, $\tilde\sigma_i$, are plotted in figure  \ref{fig:growth_b1_l0}a as a function of $|\tilde{n}|$
and $|\tilde{m}|$. We observe that the region in
wavenumber space defined roughly by $0<|\tilde{n}|<1/2,\mbox{~and~}1/2<|\tilde{m}|<1$ is unstable, with the maximum growth rate
occurring for zonal structures ($\tilde{n}=0$) with $|\tilde{m}|\simeq 0.8$. The frequency of the unstable modes is in general
non-negative ($\tilde{\sigma}_i\geq 0$) and is equal to zero only for zonal jet perturbations  ($\tilde n=0$).
Using the symmetries
(\ref{eq:symmet}), this implies that  real unstable  mean flow perturbations $\delta Z$  propagate in the retrograde direction
if $\tilde n \ne 0$ and
are stationary when $\tilde n=0$.
As $\tilde \varepsilon$ increases the instability region expands roughly covering the sector $1/2<|\tilde{N}|<1$ and
a second instability branch with smaller growth rates appears for $|\tilde{N}|>1$. This is illustrated in figure
\ref{fig:growth_b1_l0}b showing the growth rate for $\tilde \varepsilon=10\tilde \varepsilon_c$ (marked as \ref{fig:growth_b1_l0}b
in figure \ref{fig:emin}). Note also that for both values of the energy input rate, the zonal structures have a larger growth rate
than the non-zonal structures, a result that holds for any $\tilde \varepsilon$ when
$\tilde \beta < \tilde \beta_{min}$.


\begin{figure}
\centerline{\includegraphics[width=5in]{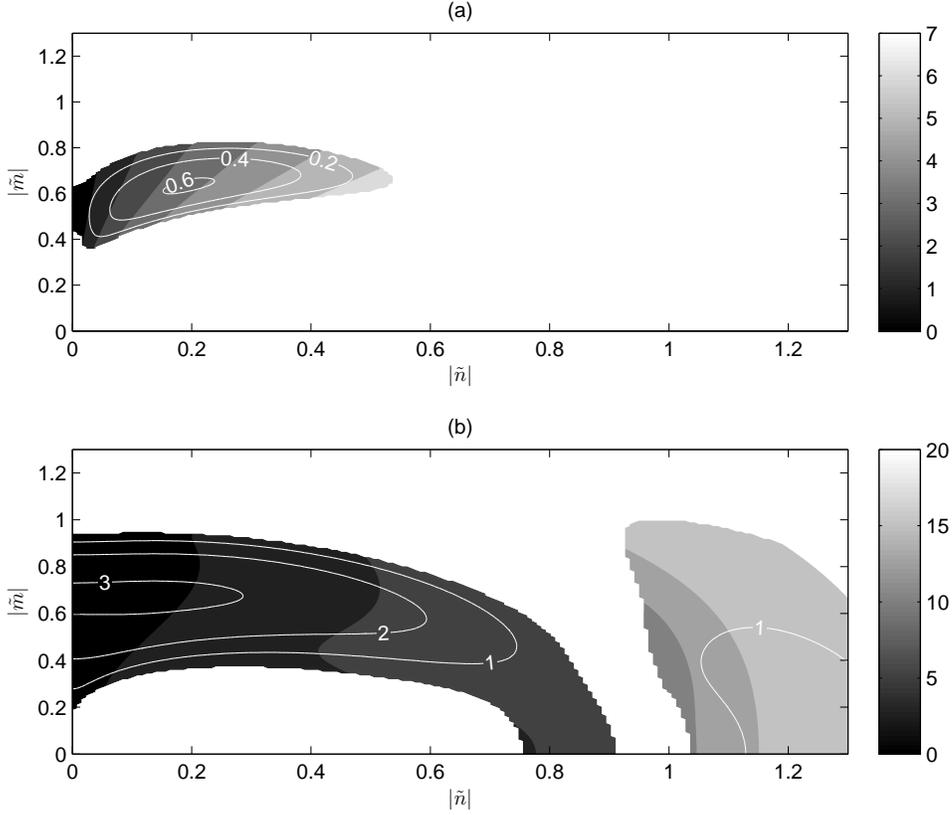}}
\caption{The same as in figure  \ref{fig:growth_b1_l0}, but for $\tilde{\beta}=10$.}
\label{fig:growth_b10_l0}
\end{figure}

\begin{figure}
\centerline{\includegraphics[width=5in]{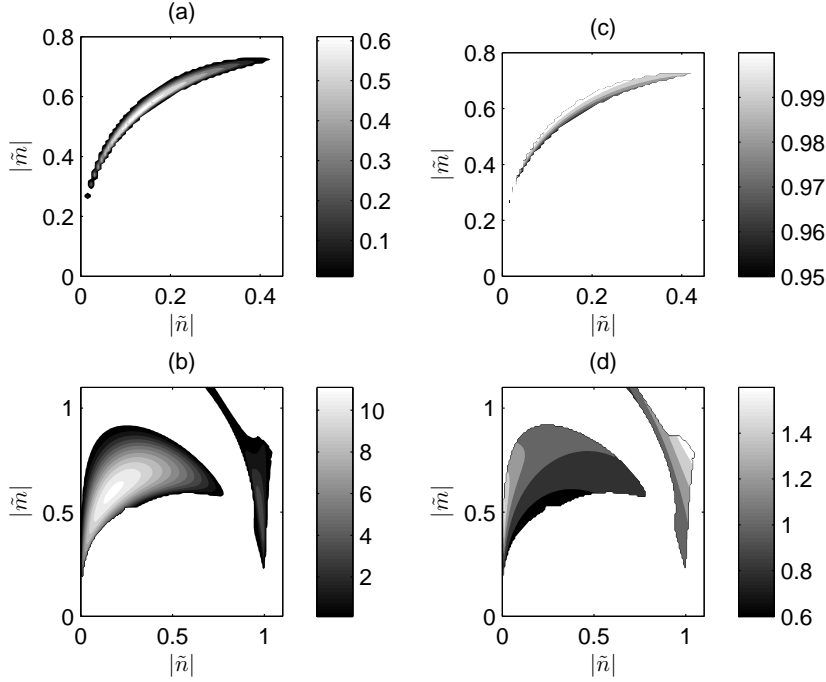}}
\caption{Dispersion relation of the unstable modes for $\tilde{\beta}=100$. Growth rate $\tilde{\sigma}_r$ as a function of
wavenumbers $(\tilde{n}, \tilde{m})$ at (a) $\tilde{\varepsilon}=2\tilde \varepsilon_c$ and (b) $\tilde{\varepsilon}=50\tilde \varepsilon_c$.
Only positive values are shown. Ratio of the frequency of the unstable modes $\tilde{\sigma}_i$ over the
corresponding frequency of a Rossby wave with the same wavenumbers $\tilde\sigma_R$ at (c) $\tilde{\varepsilon}=2\tilde \varepsilon_c$
and (d) $\tilde{\varepsilon}=50\tilde \varepsilon_c$. Values of one denote an exact match with the Rossby wave phase speed.}
\label{fig:growth_b100_l0}
\end{figure}

For $\tilde \beta > \tilde \beta_{min}$ the non-zonal structures have always larger growth rate. This is illustrated in
figures  \ref{fig:growth_b10_l0} and \ref{fig:growth_b100_l0}, showing the growth rates and frequencies of the unstable
modes for $\tilde{\beta}=10$ and  $\tilde{\beta}=100$ respectively.   For larger  $\tilde \beta$ values there is tendency for the
frequency of the unstable modes to conform to the corresponding Rossby wave frequency
\begin{equation}
\tilde \sigma_R=\frac{\tilde{\beta}\tilde{n}}{\tilde{n}^2+\tilde{m}^2} ~ ,
\end{equation}
a tendency that does not occur for smaller $\tilde \beta$. A comparison between the frequency of the unstable mode and
the Rossby wave frequency  is  shown in figures \ref{fig:growth_b100_l0}(c),(d) in a plot
of $\tilde{\sigma}_i/\tilde \sigma_R$.  For slightly supercritical
$\tilde \varepsilon$, the ratio is close to one and the unstable modes satisfy the Rossby wave dispersion relation. At higher
supercriticalities though, $\tilde{\sigma}_i$ departs from the Rossby wave frequency (by as much as $40 \%$ for the
case of $\tilde \varepsilon =50 \tilde \varepsilon_c$ shown in figure \ref{fig:growth_b100_l0}(d)).

%

\section{Equilibration of the S3T instabilities}

\begin{figure}
\centerline{\includegraphics[width=5in]{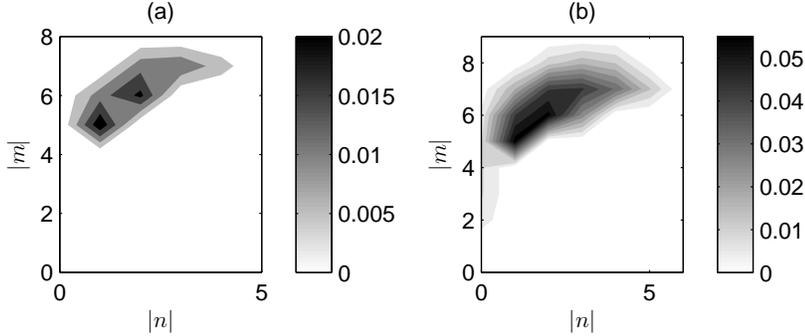}}
\vspace*{-50mm}
\caption{The growth rate, $\sigma_r$ as a function of the wavenumbers $|n|$, $|m|$ of the coherent structures at (a)
$\varepsilon/ \varepsilon_c=4$ and (b) $\varepsilon/ \varepsilon_c=10$  (only positive values of $\sigma_r$ are shown).
The growth rates are calculated for structures in the doubly periodic channel considered in the non-linear simulations
and in the presence of hyper-diffusion with coefficient $\nu=1.19\cdot 10^{-6}$.}
\label{fig:growth}
\end{figure}

\begin{figure}
\noindent\includegraphics[width=5in]{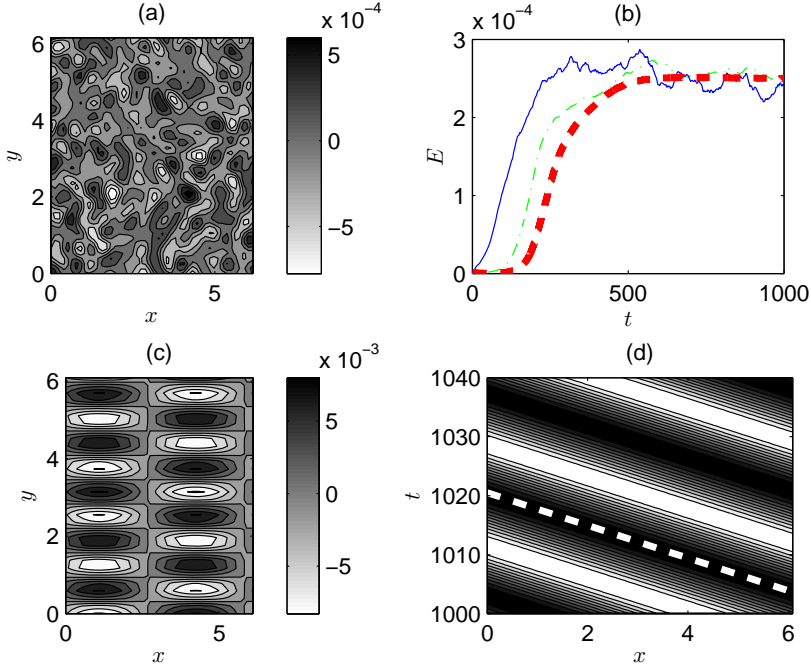}\\
\vspace*{-5mm}
\caption{Equilibration of the S3T instabilities. (a) Streamfunction of the initial perturbation. (b) Energy
evolution of the initial perturbation shown in panel (a) as obtained from the integration of the S3T equations (\ref{eq:Q_evo2})
and (\ref{eq:cov_evo2}) (dashed line) and from the integration of the ensemble quasi-linear (EQL)
system (\ref{eq:q_evo})-(\ref{eq:Q_evo}) with $N_{ens}=10$ (solid line) and $N_{ens}=100$ (dash-dotted line) ensemble members
that is discussed in section 6. (c) Snapshot of the streamfunction $\Psi_{eq}$
of the traveling wave structure and (d) Hovm\"oller diagram of $\Psi_{eq}(x,y=\upi/4,t)$ for the finite equilibrated traveling wave.
The thick dashed line shows the phase speed obtained from the stability equation (\ref{eq:dispersion}). The energy input rate
is $\tilde{\varepsilon}=4\varepsilon_c$ and $\tilde{\beta}=100$.}
\label{fig:equil1}
\end{figure}

We now investigate the equilibration of the instabilities by the discretized S3T system   (\ref{eq:Q_evo2}), (\ref{eq:cov_evo2}) in
a doubly periodic channel of size $2\upi \times 2 \upi$. We consider the parameter values
chosen in the non-linear simulations discussed in section 2 ($\beta=10$, $r=0.01$, $\nu=1.19\cdot 10^{-6}$, $K_f=10$
and $\Delta K_f=1$). For these parameters, $\tilde{\beta}=100$ and therefore the integration is in the parameter region
of figure \ref{fig:emin} in which  the  non-zonal structures are  more unstable than the zonal jets. We first consider the energy
input rate $\tilde \varepsilon=4\tilde \varepsilon_c$ which corresponds to the first  case presented in section 2
\footnote{Note that $\varepsilon_c$ here refers
to the critical energy input rate when hyperdiffusion is taken into account, which is four times greater
than the critical input rate with $\nu=0$.}. The growth rates of the coherent structures for
integer values of the wavenumbers,  $n$ and $m$  are  calculated from the discrete version of equation (\ref{eq:dispersion}),
because of the  $2 \upi$  periodicity  of the channel, with  the addition  of a  hyperdiffusive dissipation term in equation
(\ref{eq:dispersion}). The resulting growth rates for this energy input rate are shown in figure \ref{fig:growth}(a). For these
parameters the zonal jet perturbations are stable and are not expected to emerge, while a large number of non-zonal modes are
unstable with the perturbation $(n, m)=(1, 5)$ growing the most. At $t=0$, we introduce a small random perturbation, whose
streamfunction is shown in figure \ref{fig:equil1}(a). After about $t=40/\sigma_{(1,5)}$, where $\sigma_{(1, 5)}$ is the
growth rate of $(n, m)=(1, 5)$, a checkerboard perturbation of the form $Z=\cos(x)\cos(5y)$ dominates the large scale flow.
The evolution of the energy of the large scale
flow that is shown in figure \ref{fig:equil1}(b) increases rapidly and eventually saturates after about $t=150/ \sigma_{(1, 5)}$.
At this point the large scale flow gets attracted to a traveling wave finite amplitude equilibrium structure close in form to
the harmonic $Z=\cos(x)\cos(5y)$ (cf. figure \ref{fig:equil1}(c)), drifting westward. The Hovm\"oller diagram of
$\psi(x, y=\upi/4, t)$, shown in \ref{fig:equil1}(d), illustrates that the phase speed of the
traveling wave is approximately equal to the phase speed of the unstable $(n, m)=(1, 5)$ eigenmode
that is also indicated in the figure.

\begin{figure}
\noindent\includegraphics[width=5in]{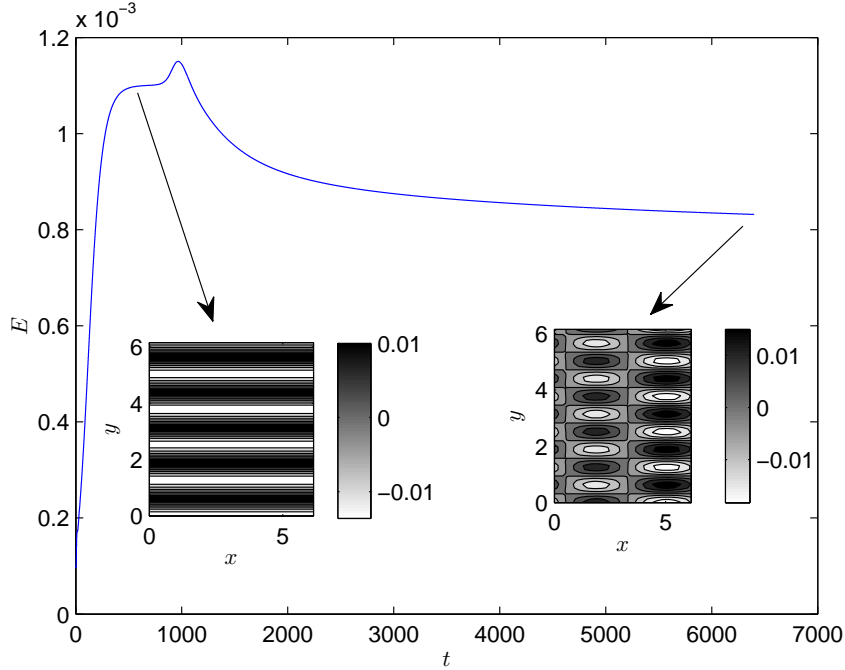}\\
\vspace*{-5mm}
\caption{Energy evolution of an initial jet perturbation $\delta Z=0.001\cos(5y)$ for $\tilde{\varepsilon}=10\varepsilon_c$ and
$\tilde{\beta}=100$. The insets show a snapshot of the mean flow streamfunction at $t=500$ (left) and the streamfunction of
the equilibrated structure at $t=6500$ (right).}
\label{fig:equil2}
\end{figure}

Consider now the energy input rate $\tilde \varepsilon=10\tilde \varepsilon_c$ for which the growth rates are shown in
figure \ref{fig:growth}(b). While the maximum growth rate still occurs for the $(|n|, |m|)= (1, 5)$ non-zonal structure,
zonal jet perturbations are unstable as well. In previous studies of S3T dynamics restricted to
the interaction between zonal flows and turbulence, these initially S3T unstable jet structures were found to equilibrate
at finite amplitude. However, in the context of the generalized S3T analysis
in this work that takes into account the dynamics of the interaction between coherent non-zonal structures and jets, we find
that these S3T jet equilibria can be saddles: stable to zonal jet perturbations but unstable to non-zonal perturbations. To show
this, we consider the evolution of a small zonal jet perturbation $\delta Z=0.001\cos(5y)$ that is shown in figure \ref{fig:equil2}.
The initial perturbation grows exponentially and its energy saturates at about $t=500$ (a snapshot of the streamfunction
at this time is shown at the left inset in figure \ref{fig:equil2}). But soon after, non-zonal undulations appear and start to
grow and the flow transitions to the stable $Z=\cos(x)\cos(5y)$ traveling wave state that is also shown in figure
\ref{fig:equil2}. As a result, the finite equilibrium zonal jet structure is S3T unstable to coherent non-zonal
perturbations and is not expected to appear in non-linear simulations despite the fact that the zero flow equilibrium is
unstable to zonal jet perturbations. We will elaborate more on this issue in the next section.

\begin{figure}
\noindent\includegraphics[width=5in]{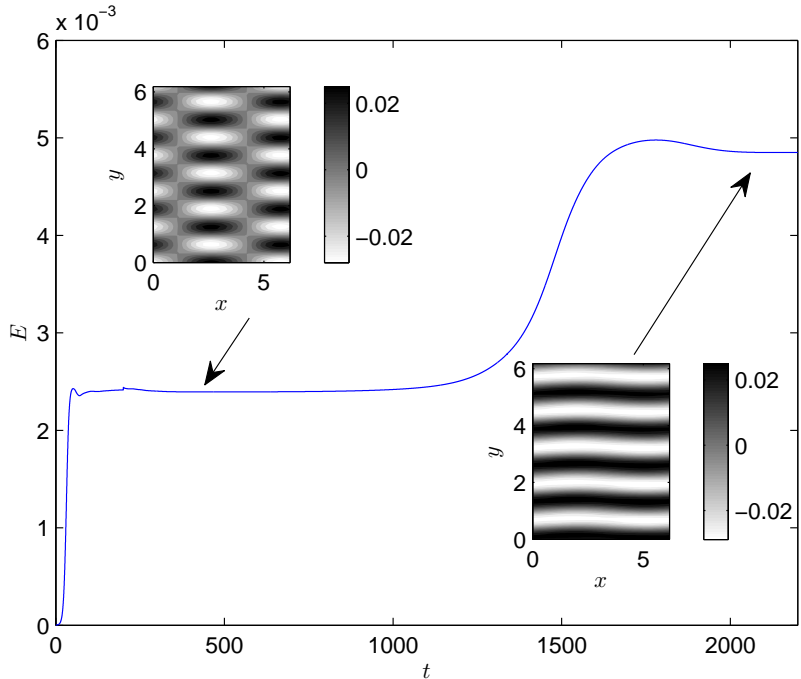}\\
\vspace*{-5mm}
\caption{Energy evolution of the initial non-zonal perturbation $\delta Z=0.01\cos(x)\cos(5y)$ for $\tilde{\varepsilon}=30\varepsilon_c$ and
$\tilde{\beta}=100$. The insets show a snapshot of the mean flow streamfunction at $t=200$ (left) and the streamfunction of
the equilibrated structure at $t=2200$ (right).}
\label{fig:equil3}
\end{figure}

\begin{figure}
\noindent\includegraphics[width=5in]{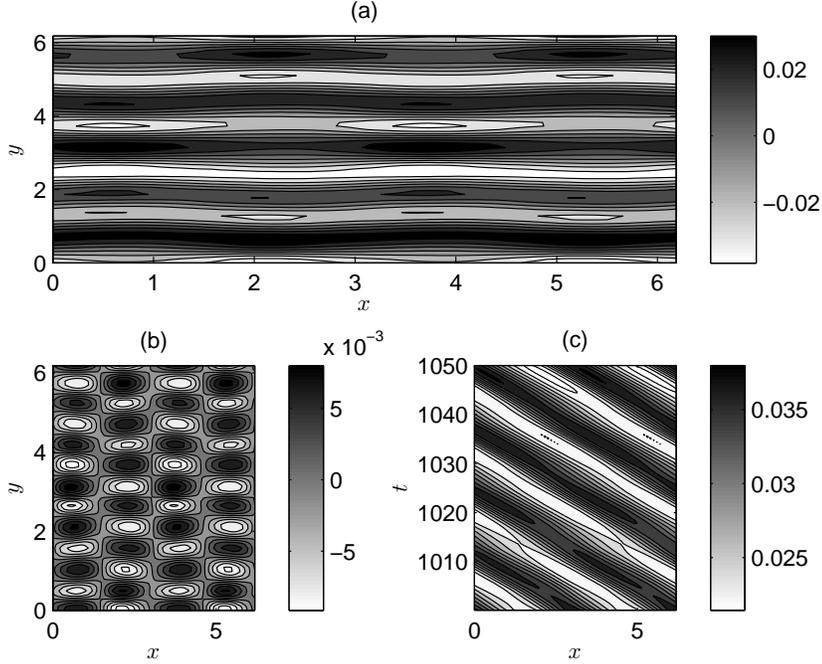}\\
\vspace*{-5mm}
\caption{Mixed zonal jet-traveling wave S3T equilibrium for $\tilde{\varepsilon}=30\varepsilon_c$ and $\tilde{\beta}=100$.
(a) Snapshot of the streamfunction $\Psi_{eq}$ of the equilibrium state. (b) Contour plot of the non-zonal component
$\Psi_{eq}-\overline{\Psi_{eq}}$ of the equilibrium structure, where the overline denotes a zonal average. (c)
Hovm\"oller diagram of $\Psi_{eq}(x,y=\upi/4,t)$ for the equilibrated structure. In this mixed S3T equilibrium zonal jets
coexist with vortices reminiscent of the coherent structures observed in Jupiter.}
\label{fig:equil4}
\end{figure}

Finally, consider the case $\varepsilon=30\varepsilon_c$. At this energy input rate, the fast growing non-zonal perturbations
cannot equilibrate, as the finite amplitude non-zonal traveling wave equilibria become S3T unstable. This is
illustrated in figure \ref{fig:equil3} showing the evolution of a small non-zonal perturbation $\delta Z=0.01\cos(x)\cos(5y)$.
After the saturation of the initial instability at about $t=200$, the flow transitions slowly from the traveling wave
$Z=\cos(x)\cos(5y)$ state shown at the left inset in figure \ref{fig:equil3} to the jet equilibrium state shown at the right
inset in figure \ref{fig:equil3}. Note however, that the jet equilibrium structure is not zonally symmetric. This is
a new type of S3T equilibrium: it is a mix between a zonal jet and a non-zonal traveling wave and actually S3T analysis 
reveals multiple mixed state equilibria. This is clearly illustrated 
in figure \ref{fig:equil4} showing the structure of a different equilibrium state for the same parameters. The equilibrium structure
consists of a large amplitude zonally symmetric jet with small amplitude non-zonal
propagating vortices embedded in it. These vortices that are shown in figure \ref{fig:equil4}(b) to have approximately the compact
support structure $\Psi=\cos(2x)\cos(6y)$ propagate westward as shown in the Hovm\"oller diagram in figure \ref{fig:equil4}(c).

\section{Comparison to non-linear simulations}

Within the context of the second order cumulant closure, the S3T formulation allows the identification of statistical
turbulent equilibria. These S3T equilibria and their stability properties are manifest even in
single realizations of the turbulent system. For example, previous studies using S3T obtained zonal jet equilibria in barotropic, shallow water and baroclinic flows in close correspondence with
observed jets in planetary flows \citep{FI-07,FI-08,Farrell-Ioannou-2009-closure,Farrell-Ioannou-2009-equatorial}. In
addition, previous studies of S3T dynamics restricted to the interaction between zonal flows and turbulence in a $\beta$-plane
channel showed that when the energy input rate is such that the zero mean flow equilibrium is unstable,
zonal jets also appear in the non-linear simulations with the structure (scale
and amplitude) predicted by S3T \citep{Srinivasan-Young-12,Constantinou-etal-2012}.

A very useful intermediate model that retains the wave-mean flow dynamics of the S3T system while relaxing the infinite
ensemble approximation can be constructed by ignoring in (\ref{eq:q_evo}) the non-linear term $f^{nl}$. Then
(\ref{eq:q_evo})-(\ref{eq:Q_evo}) become an ensemble quasi-linear system (EQL) in which the ensemble mean can be calculated
from a finite number of ensemble members. Its integration is done as follows. Using the same pseudo-spectral code as in the non-linear
simulations, $N_{ens}$ separate integrations of (\ref{eq:q_evo}) are performed at each time step with the eddies evolving
according to the instantaneous flow. Then the ensemble average vorticity flux divergence is calculated as the average over
the $N_{ens}$ simulations and (\ref{eq:Q_evo})
is stepped forward in time according to those fluxes. The S3T equilibria manifest in the EQL integrations with
the addition of some 'thermal noise' due to the stochasticity of the forcing that is retained in this system. This is
shown in figure \ref{fig:equil1}b where the energy growth of the coherent structure for $N_{ens}=10$ and $N_{ens}=100$ is
plotted for the same
parameters as the S3T integration. We observe that the energy of the coherent structure in the EQL
integrations fluctuates around the values predicted by the S3T system with the fluctuations decreasing as
$1/\sqrt{N_{ens}}$. However, even with only 10 ensemble members we get an estimate that
is very close to the theoretical estimate of the infinite ensemble members obtained from the S3T integration. The
traveling wave equilibrium and its phase speed in the quasi-linear integrations are also in very good agreement
with the corresponding structure and phase speed obtained from the S3T integration (not shown). Since the EQL system
accurately captures the characteristics of the emerging structures and it is much faster to integrate, we will use
it to test the predictions of S3T for the emergence and equilibration of zonal and non-zonal coherent structures in
non-linear simulations. We will therefore present comparisons of the integrations of the EQL system
with $N_{ens}=10$ with single realizations of the non-linear equations.

\begin{figure}
\centerline{\includegraphics[width=5in]{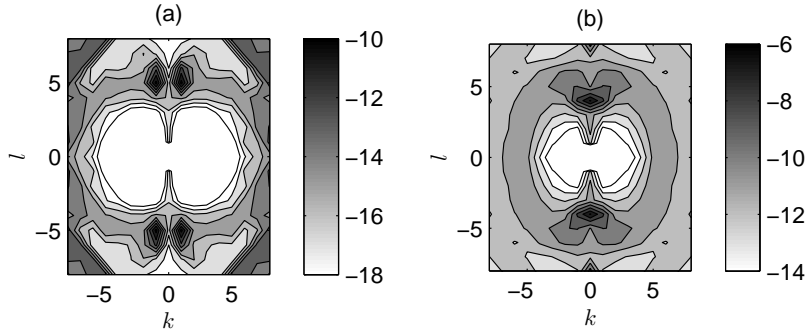}}
\vspace*{-45mm}
\caption{Time averaged energy power spectra, $\log(\hat{E}(k,l))$, obtained from the ensemble
quasi-linear (EQL) simulations at  (a) $\varepsilon/ \varepsilon_c=4$ and (b) $\varepsilon/ \varepsilon_c=50$ or
$\varepsilon=3.3\varepsilon_{nl}$. There is a very good agreement with the corresponding
spectra obtained from the non-linear (NL) simulations (cf. figures \ref{fig:NL_snap1}-\ref{fig:NL_snap2}).}
\label{fig:QL_spec}
\end{figure}

For the parameters chosen ($\tilde\beta=100$), S3T predicts emergence of non-zonal structures when
the energy input rate exceeds the critical threshold $\tilde\varepsilon_c$ and equilibration of the incipient instabilities
into finite amplitude structures that should manifest in the non-linear simulations. The rapid increase of the nzmf index in
the non-linear (NL) and quasi-linear (EQL) simulations for $\varepsilon>\varepsilon_c$ shown in figure \ref{fig:zmf}, illustrates that this
regime transition in the flow is accurately predicted by S3T and that the quasi-linear and non-linear dynamics share the
same bifurcation structure. In addition, the stable S3T equilibria are in principle viable repositories of energy in the turbulent
flow and the non-linear system is expected to visit their attractors for finite time intervals. Indeed for
$\varepsilon=4\varepsilon_c$ the traveling wave equilibrium with $(|k|, |l|)=(1, 5)$ that emerges out of random initial conditions,
is the dominant structure in the NL simulations. Comparison of the energy spectra obtained from the EQL and the
NL simulations shown in figures \ref{fig:QL_spec}a and \ref{fig:NL_snap1}a respectively, reveals that the amplitude of
this structure in the quasi-linear and in the non-linear dynamics almost matches. Remarkably, the phase speed of the S3T
traveling wave matches with the corresponding phase speed of the $(|k|, |l|)=(1, 5)$ structure observed in
the non-linear simulations, as can be seen in the Hovm\"oller diagram in figure \ref{fig:NL_snap1}(b). Such an agreement in the
characteristics of the emerging structures between the EQL and NL simulations occurs for a wide range of
energy input rates as can be seen by comparing the nzmf indices in figure \ref{fig:zmf}. As a result, S3T predicts
the dominant non-zonal propagating structures in the non-linear simulations, as well as their amplitude and phase speed.

The second transition in which zonal jets emerge is more intriguing. The stability equation
(\ref{eq:dispersion}) predicts that the zonal structures become S3T unstable at $\varepsilon_{sz}=5.2\varepsilon_c$. As discussed
in the previous section, the finite amplitude zonal jet equilibria are structurally unstable and the flow stays on
the attractor of the non-zonal
traveling wave equilibria (cf. figure \ref{fig:equil2}). We determined that the
jet equilibria are structurally unstable for $\varepsilon_{sz}<\varepsilon<\varepsilon_{nl}$. When
$\varepsilon>\varepsilon_{nl}$, the non-zonal traveling wave equilibria become S3T unstable
while at these parameter values the S3T system has mixed zonal jet-traveling wave
equilibria which are stable (cf. figure \ref{fig:equil4}). In both NL and EQL simulations, similar mixed
zonal flow-traveling wave structures are evident (cf. figures \ref{fig:NL_snap2}, \ref{fig:QL_spec}).
However, there is a small discrepancy regarding the second transition between the EQL and NL simulations, as
the energy input rate threshold for the emergence of jets is slightly larger in the NL simulations compared
to the corresponding EQL threshold (cf. figure 1). This discrepancy possibly occurs due to the fact that the
exchange of instabilities between the
mixed jet-traveling wave equilibria and the pure traveling wave equilibria depends on the equilibrium structure
$[Z^E, C^E]$. Small changes for example in $C^E$ that might be caused by the eddy-eddy terms neglected in
S3T can cause the exchange of instabilities to occur at slightly different energy input rates. It was shown in a
recent study that when the effect of the eddy-eddy terms is taken into account by obtaining $C^E$ directly from the
nonlinear simulations, the S3T stability analysis performed on this corrected equilibrium states accurately predicts
the energy input rate for the emergence of jets in the nonlinear simulations \citep{Constantinou-etal-2012}. The
power spectrum obtained from the EQL simulations for $\varepsilon=3.3\varepsilon_{nl}$ (cf. figure \ref{fig:QL_spec}b)
shows that both the scale and the amplitude of the zonal jets is captured by S3T. Such an agreement again holds for
a wide range of energy input rates, as the zmf indices obtained from the EQL and the NL simulations indicate.
In summary, S3T predicts the characteristics of both non-zonal propagating structures and of zonal jets in the
non-linear simulations.

\section{Discussion-conclusions}

A theory for the emergence of zonal jets and non-zonal coherent structures in barotropic beta-plane
turbulence is presented in this work. This is one of the simplest
models that retains the relevant dynamics of self-organization of turbulence into large scale coherent
structures and is a standard and extensively studied testbed for theories regarding the emergence and
maintenance of zonal jets and coherent structures in planetary flows.


Non-linear simulations of a stochastically forced barotropic beta-plane channel show two major flow
transitions as the energy input rate of the forcing increases. In the first, the translational symmetry
in the flow is broken in both directions with the emergence of propagating coherent non-zonal waves that approximately follow
the Rossby wave dispersion. The power in these non-zonal structures increases with the energy input rate
until the second transition occurs with the emergence of robust zonal jets. Although after the second transition the
zonal jets contain over half the energy in the flow, there is significant power in the non-zonal structures, which
remain coherent and propagate in the retrograde direction with phase speeds that do not satisfy the Rossby wave
dispersion.

The two flow transitions and the characteristics of both the non-zonal structures and the zonal jets
are then investigated using a proper generalization of Stochastic Structural Stability Theory (S3T). In S3T,
the turbulent flow dynamics and statistics are expressed as a systematic cumulant expansion which is truncated
at second order. With the interpretation of the ensemble average as a Reynolds average
over the fast turbulent eddies adopted in this work, the second order cumulant expansion results in a closed,
non-linear dynamical system that governs the joint evolution of slowly varying, spatially localized
coherent structures with the second order statistics of the rapidly evolving turbulent eddies.
The fixed points of this autonomous, deterministic non-linear system define statistical equilibria,
the stability of which are amenable to the usual treatment of linear and non-linear stability analysis.

The linear stability of the homogeneous S3T equilibrium with no mean velocity was examined analytically. Structural
instability was found to occur for perturbations with smaller scale
than the forcing, when the energy input rate $\tilde\varepsilon=\varepsilon K_f^2/r^3$ is larger than a certain threshold
$\tilde\varepsilon_c$ that depends on $\tilde\beta=\beta /(rK_f)$. It was found that when $\tilde\beta$ is small or order
one, both zonal jets and non-zonal structures are unstable when the energy input rate is larger than $\tilde\varepsilon_c$,
with the maximum growth rate occurring for stationary zonal structures. When
$\tilde\beta$ is large, non-zonal structures first become unstable as the input rate increases past $\tilde\varepsilon_c$
with zonal jet structures becoming unstable at larger energy input rates. The maximum growth rate occurs in this case
for non-zonal structures that propagate in the retrograde direction. These waves follow the Rossby wave dispersion
for low supercritical values of the energy input rate, but propagate with different phase speeds at higher
supercriticality. The equilibration of the unstable, exponentially growing coherent structures for large $\tilde\beta$ was then
studied through numerical integrations of the S3T dynamical system. When the forcing amplitude is slightly
supercritical, the finite amplitude traveling wave equilibrium has a structure close
to the corresponding unstable non-zonal perturbation with the same scale. When the forcing amplitude is highly
supercritical, the instabilities equilibrate to mixed states consisting of strong zonal jets with
smaller amplitude traveling waves embedded in them.

The predictions of S3T were then compared to the results that obtain in the non-linear simulations. The critical
threshold above which coherent non-zonal structures are unstable according to the stability analysis of the
S3T system was found to be in excellent agreement with the critical value above which non-zonal structures acquire
significant power in the non-linear simulations. The scale, phase speed and amplitude of the dominant structures
in the non-linear simulations were also found to correspond to the structures predicted by S3T. In addition, the threshold for
the emergence of jets, which is identified in S3T as the energy input rate at which an S3T stable, finite amplitude
zonal jet equilibrium exists, was found to roughly match the corresponding threshold for jet formation in the non-linear
simulations, with the emerging jet scale and amplitude being accurately obtained using S3T.

In summary, S3T predicts the two regime transitions in the turbulent flow
as the energy input rate is increased: the emergence of
coherent, propagating non-zonal structures and the emergence of zonal jets. It also predicts the characteristics of
the emerging structures (their scales and their phase speed), as well as their amplitude. These results provide a
concrete example that
large scale structure in barotropic turbulence, whether it is zonal jets or non-zonal coherent structures, can arise from
systematic self-organization of the turbulent Reynolds stresses by spectrally non-local interactions and in the absence
of a turbulent cascade. The analysis reveals that the coherent structures emerge as unstable modes of the homogeneous
statistical equilibrium. This instability shares the universal properties of pattern formation.
In this work we have shown that the emergence of striped patterns (zonal jets) is preceded by the emergence
of oscillating patterns (propagating waves). The analogy with pattern formation and the universality of the
underlying dynamics may prove fruitful for understanding the domain of attraction of the non-zonal equilibria,
as was previously done for the case of convection \citep{Busse-78}. This is part of ongoing
research efforts by the authors and will be reported in the future.

Finally, we note that some of the characteristics of the coherent structures found in the
barotropic beta-plane model may not accurately reproduce  the characteristics of observed structures in
the atmosphere or ocean. This should be no surprise, since the barotropic model lacks some of the important physics
(baroclinicity, etc). For example, the oceanic vortex rings do not resemble the same plane wave or compact support
structure of the coherent structures reported in this study. However, similarly with the structures that form under
S3T dynamics, the vortex rings share the characteristic that they
act as a long-lived repository of energy in the turbulent flow. Therefore their connection to the reported coherent
structures needs to be further elucidated and is an attractive venue for future research.

This research was supported by the EU FP-7 under the PIRG03-GA-2008-230958 Marie Curie Grant. The authors acknowledge
the hospitality of the Aspen Center for Physics supported by the NSF (under grant No. 1066293), where part of this
work was done. The authors would also like to thank Navid Constantinou, Brian Farrell, James Cho, Freddy Bouchet and
Brad Marston for fruitful discussions.

\appendix
\section{Physical parameters for the Earth's atmosphere and ocean and for the Jovian atmosphere}\label{appB}

In this Appendix we discuss the relevant physical parameters (the forcing length scale, the dissipation
time scale and the values of $\beta$) for the Earth's atmosphere and ocean and for the Jovian atmosphere. For the Earth's
midlatitude atmosphere ($\beta=2\cdot 10^{-11}~\mbox{m}^{-1}\mbox{s}^{-1}$), we assume that the energy is injected
at the cyclone scale of $1/K_f=1000~\mbox{km}$ and that
the eddy dissipation time scale at midlatitudes is $1/r = 10~\mbox{days}$. In addition, an energy transfer from the
mean to the eddies of the order of $1.3~\mbox{Wm}^{-2}$ is typically found in studies of the Lorenz
cycle in the atmosphere, while there is also another $20~\mbox{Wm}^{-2}$ available through diabatic heating
\citep{Peixoto-Oort-1992}. Assuming that a small fraction of the order of 5\% is transferred into large scale eddies,
we obtain a total amount of $2.3~\mbox{Wm}^{-2}$, which when injected over the troposphere with a scale height of 8 km,
corresponds to an energy injection rate $\varepsilon=3\cdot~10^{-4}~\mbox{m}^2\mbox{s}^{-3}$. For the Earth's ocean, we assume that the
eddy energy is injected at the deformation scale $1/K_f=20~\mbox{km}$, while we consider that the eddy dissipation
time scale is $1/r=1000~\mbox{days}$ \citep{Berloff-09a} and the energy injection rate is
$\varepsilon=10^{-9}~\mbox{m}^2\mbox{s}^{-3}$ \citep{Sukoriansky-etal-2007}. For the Jovian atmosphere
at midlatitudes ($\beta=2.5\cdot 10^{-12}~\mbox{m}^{-1}\mbox{s}^{-1}$), we assume
that energy is injected at the scale of convective storms $1/K_f=100~\mbox{km}$ with a rate
$\varepsilon=0.5\cdot 10^{-5}~\mbox{m}^2\mbox{s}^{-3}$ \citep{Galperin-etal-13}. Since the eddy dissipation rate is
not known, we obtain an estimate by assuming that the observed root mean square velocity fluctuations $U_{rms}$ satisfy
$U_{rms}^2=\varepsilon/r$. Taking $U_{rms}=50~\mbox{ms}^{-1}$ \citep{Galperin-etal-13}, we obtain $1/r=5800~\mbox{days}$.
It should be noted that these values are indicative order of magnitude estimates.

\section{Calculation of the dispersion relation and its properties}\label{appA}

In this Appendix we derive the dispersion relation (\ref{eq:dispersion}), which determines the stability of
zonal as well as non-zonal perturbations in homogeneous turbulence.
We follow closely the treatment of
\cite{Srinivasan-Young-12}. We first rewrite (\ref{eq:Q_evo2}), (\ref{eq:cov_evo2}) in terms of the variables
$\tilde{x}=x_1-x_2$, $\overline{x}=(1/2)(x_1+x_2)$, $\tilde{y}=y_1-y_2$ and $\overline{y}=(1/2)(y_1+y_2)$. The
derivatives transform into this new system of coordinates to $\partial_{x_i}=(1/2)\partial_{\overline{x}}+
(-1)^{i+1}\partial_{\tilde{x}}$, $\partial_{y_i}=(1/2)\partial_{\overline{y}}+(-1)^{i+1}\partial_{\tilde{y}}$, $\Delta_i=\tilde{\Delta}+(1/4)\overline{\Delta}+(-1)^{i+1}\partial_{\tilde{y}\overline{y}}^2+
(-1)^{i+1}\partial_{\tilde{x}\overline{x}}^2$, with $\tilde{\Delta}=\partial_{\tilde{x}\tilde{x}}^2+
\partial_{\tilde{y}\tilde{y}}^2$ and $\overline{\Delta}=\partial_{\overline{x}\overline{x}}^2+
\partial_{\overline{y}\overline{y}}^2$. It
is also convenient to introduce the streamfunction covariance
$S(\tilde{x}, \overline{x}, \tilde{y}, \overline{y})\equiv\left<\psi_1'\psi_2'\right>$, which is related to
$C(\tilde{x}, \overline{x}, \tilde{y}, \overline{y})$ via:
\begin{equation}
C=\left<\zeta_1'\zeta_2'\right>=\left<\Delta_1\psi_1'\Delta_2\psi_2'\right>=
\Delta_1\Delta_2 S=\left[\left({\tilde{\Delta}}+{1\over 4}\overline{\Delta}\right)^2-\left(\partial_{\tilde{x}\overline{x}}^2+\partial_{\tilde{y}\overline{y}}^2\right)^2\right] S.
\end{equation}
Equations (\ref{eq:Q_evo2}), (\ref{eq:cov_evo2}) then become in the absence of hyper-viscosity ($\nu=0$):
\begin{eqnarray}
& &\left[\partial_t+\overline{U}\partial_{\overline{x}}+\tilde{U}\partial_{\tilde{x}}+\overline{V}
\partial_{\overline{y}}+\tilde{V}\partial_{\tilde{y}}\right]C+\left[(\beta+\overline{Z}_y)\partial_{\overline{x}}+
\tilde{Z}_y\partial_{\tilde{x}}-\overline{Z}_x\partial_{\overline{y}}-\tilde{Z}_x\partial_{\tilde{y}}\right]
\left(\tilde{\Delta}+\frac{1}{4}\overline{\Delta}\right)S\nonumber\\ &-&\left[2(\beta+\overline{Z}_y)\partial_{\tilde{x}}+
\frac{1}{2}\tilde{Z}_y\partial_{\overline{x}}-2\overline{Z}_x\partial_{\tilde{y}}-\frac{1}{2}
\tilde{Z}_x\partial_{\overline{y}}\right]
\left(\partial_{\tilde{x}\overline{x}}^2+\partial_{\tilde{y}\overline{y}}^2\right)S=-2rC+\Xi,
\label{eq:dcdt_y}
\end{eqnarray}
\begin{equation}
\partial_tZ+U\partial_xZ+V(\beta+\partial_yZ)=(\partial_{\tilde{x}\overline{y}}^2-\partial_{\tilde{y}\overline{x}}^2)(\partial_{\tilde{x}\overline{x}}^2+
\partial_{\tilde{y}\overline{y}}^2) S|_{\tilde{x}=\tilde{y}=0}-rZ,\label{eq:dudt_y}
\end{equation}
where $(\overline{U}, \overline{V})=(1/2)(U_1+U_2, V_1+V_2)$, $(\tilde{U}, \tilde{V})=(U_1-U_2, V_1-V_2)$,
$(\overline{Z}_x, \overline{Z}_y)=(1/2)(\partial_{x_1}+\partial_{x_2}, \partial_{y_1}+\partial_{y_2})Z$ and
$(\tilde{Z}_x, \tilde{Z}_y)=(\partial_{x_1}-\partial_{x_2}, \partial_{y_1}-\partial_{y_2})Z$.

The forcing covariance $\Xi$ is homogeneous and as a result it depends only on the difference coordinates, $\tilde x$ and $\tilde y$. It
can then be readily shown from (\ref{eq:dcdt_y})-(\ref{eq:dudt_y}), that the state with no coherent flow ($U^E=V^E=Z^E=0$) and with the
homogeneous vorticity covariance $C^E(\tilde{x}, \tilde{y})=\Xi /(2r)$ (implying also that the streamfunction covariance $S^E$ is homogenous)
is a fixed point of the S3T system. The stability of this homogeneous equilibrium, can be addressed by first linearizing the
S3T system about it:
\begin{eqnarray}
\partial_t\delta C&=&-\left(\delta \tilde{U}\partial_{\tilde{x}}+\delta\tilde{V}\partial_{\tilde{y}}
\right)C^E-\left(\delta\tilde{Z}_y\partial_{\tilde{x}}-\delta \tilde{Z}_x
\partial_{\tilde{y}}\right)\tilde{\Delta}S^E\nonumber\\ &-&\beta
\left\{\left[\tilde{\Delta}+\frac{1}{4}\overline{\Delta}\right]\partial_{\overline{x}}-
2(\partial_{\tilde{x}\overline{x}}^2+\partial_{\tilde{y}\overline{y}}^2)\partial_{\tilde{x}}
\right\}\delta S-2r\delta C,\label{eq:Lc}
\end{eqnarray}
\begin{equation}
\partial_t\delta Z=-\beta \delta V+(\partial_{\tilde{x}\overline{y}}^2-\partial_{\tilde{y}\overline{x}}^2)
(\partial_{\tilde{x}\overline{x}}^2+\partial_{\tilde{y}\overline{y}}^2)\delta S|_{\tilde{x}=\tilde{y}=0}
-r\delta Z,\label{eq:Lu}
\end{equation}
where $\delta Z$, $\delta\tilde{U}$, $\delta\tilde{V}$, $\delta\tilde{Z}_x$, $\delta\tilde{Z}_y$, $\delta C$ and
$\delta S$ are small perturbations in the ensemble mean vorticity, velocities and vorticity gradients and in the eddy vorticity and
streamfunction covariances respectively, and then performing an eigenanalysis of the linearized
equations (\ref{eq:Lc})-(\ref{eq:Lu}).

We consider a harmonic vorticity perturbation of the form $\delta Z=\mathrm{e}^{\mathrm{i}nx+\mathrm{i}my}
\mathrm{e}^{\sigma t}$, for which:
\begin{equation}
[\delta \tilde{U}, \delta\tilde{V},\delta\tilde{Z}_x, \delta\tilde{Z}_y]=
-2\left[\frac{m}{n^2+m^2},-\frac{n}{n^2+m^2},n,m\right]
\sin\left(\frac{n\tilde{x}}{2}+\frac{m\tilde{y}}{2}\right)
\mathrm{e}^{\mathrm{i}n\overline{x}+\mathrm{i}m\overline{y}}\mathrm{e}^{\sigma t}.\label{eq:tildeU}
\end{equation}
Taking the same form for the streamfunction covariance perturbation $\delta S = S_{nm}(\tilde{x}, \tilde{y})
\mathrm{e}^{\mathrm{i}n\overline{x}+\mathrm{i}m\overline{y}}\mathrm{e}^{\sigma t}$ and inserting it in
(\ref{eq:Lc})-(\ref{eq:Lu}) along with
(\ref{eq:tildeU}) yields:
\begin{eqnarray}
& &\left\{(\sigma+2r)\left[\left(\tilde{\Delta}-\frac{N^2}{4}\right)^2+\Delta_+^2\right]-2\mathrm{i}\beta\Delta_+
\partial_{\tilde{x}}+\mathrm{i}n\beta
\left(\tilde{\Delta}-\frac{N^2}{4}\right)\right\} S_{nm}
\nonumber\\ &=&\frac{2}{N^2}\sin\left(\frac{n\tilde{x}}{2}+\frac{m
\tilde{y}}{2}\right)\left(m\partial_{\tilde{x}}-n\partial_{\tilde{y}}\right)\left[\tilde{\Delta}^2+N^2
\tilde{\Delta}\right] S^E,\label{eq:disp1}
\end{eqnarray}
\begin{equation}
-(\sigma +r)N^2+\mathrm{i}n\beta=N^2\left(m\partial_{\tilde{x}}-n\partial_{\tilde{y}}\right)\Delta_+S_{nm}
|_{\tilde{x}=\tilde{y}=0},\label{eq:disp2}
\end{equation}
where $N^2=n^2+m^2$, $\Delta_+=n\partial_{\tilde{x}}+m
\partial_{\tilde{y}}$ and $C^E=\Xi/2r=\tilde{\Delta}^2 S^E$ is the
equilibrium vorticity covariance with zero mean flow.

Defining the Fourier transform of ${S}_{nm}(\tilde{x},\tilde{y})$ by
\begin{equation}
\hat{S}_{nm} (k,l)={1\over 2\upi} \int_{-\infty}^{\infty}\int_{-\infty}^{\infty} { S}_{nm}(\tilde{x},\tilde{y})
\mathrm{e}^{-\mathrm{i}k\tilde{x}-\mathrm{i}l\tilde{y}} \mathrm{d} \tilde{x} \mathrm{d} \tilde{y}~,
\end{equation}
we obtain from  (\ref{eq:disp1})  that the Fourier component $\hat{S}_{nm}$ satisfies:
\begin{eqnarray}
\hat{S}_{nm}&=&\frac{(mk_--nl_-)
K_-^2(K_-^2/N^2-1)\hat{S}^E(k_-,l_-)}{2\mathrm{i}\beta k(kn+ml)-\mathrm{i}n\beta (K_+^2+K_-^2)/2+
(\sigma+2r) K_+^2K_-^2}\nonumber\\ & - &\frac{(mk_+-nl_+)K_+^2(K_+^2/N^2-1)\hat{S}^E(k_+,l_+)}
{2\mathrm{i}\beta k(kn+ml)-\mathrm{i}n\beta (K_+^2+K_-^2)/2+
(\sigma+2r) K_+^2K_-^2},\label{eq:Psin}
\end{eqnarray}
with $k_\pm=k\pm n/2$, $l_\pm=l\pm m/2$, $K_\pm^2=k_\pm^2+l_\pm^2$ and
$K^2=k^2+l^2$. $\hat{S}^E=\hat{\Xi}/(2rK^4)$ is the Fourier transform
of $ S^E$, and $\hat{\Xi}$ is the Fourier transform of  $\Xi$. In addition, (\ref{eq:disp2}) becomes:

\begin{equation}
\mathrm{i}n\beta-(\sigma+r)N^2=-{N^2\over 2\upi}\int_{-\infty}^\infty\int_{-\infty}^\infty \left[nm(k^2-l^2)+(m^2-n^2)kl\right]
\hat{S}_{nm}\mathrm{d}k\mathrm{d}l=\Lambda_+-\Lambda_-,\label{eq:disper1}
\end{equation}
where
\begin{equation}
\Lambda_\pm={1\over 2\upi}\int_{-\infty}^\infty\int_{-\infty}^\infty
\frac{\left[nm(k^2-l^2)+(m^2-n^2)kl\right](mk_\pm-nl_\pm)K_\pm^2(K_\pm^2-N^2)
\hat{S}^E(k_\pm,l_\pm)}{2\mathrm{i}\beta k(kn+ml)-\mathrm{i}n\beta (K_+^2+K_-^2)/2+(\sigma+2r)
K_+^2K_-^2}\mathrm{d}k\mathrm{d}l.\label{eq:lambda}
\end{equation}

Equation (\ref{eq:disper1}) can be further simplified by noting that because the choice of $\mathbf{x}_1$ and
$\mathbf{x}_2$ is arbitrary, the forcing covariance satisfies the exchange symmetry $\Xi(x_1,x_2,y_1,y_2)=
\Xi(x_2,x_1,y_2,y_1)$. In terms of the new variables, the exchange symmetry is written as
$\Xi(\tilde{x}, \overline{x}, \tilde{y},\overline{y})=\Xi(-\tilde{x}, \overline{x},-\tilde{y},\overline{y})$, and
consequently $\hat{\Xi}$ satisfies $\hat{\Xi}(-k,-l)=\hat{\Xi}(k,l)$. As a result:
\begin{equation}
\Lambda_+=-\Lambda_-.\label{eq:lambda2}
\end{equation}
Using (\ref{eq:lambda2}) and shifting the axes in the resulting integrals ($k\rightarrow k+n/2$ and
$l\rightarrow l+m/2$), reduces (\ref{eq:disper1}) to the following dispersion relation:
\begin{eqnarray}
& &\int_{-\infty}^{\infty}\int_{-\infty}^{\infty}\frac{(mk-nl)\left[nm(k_+^2-l_+^2)+(m^2-n^2)k_+l_+
\right]K^2(K^2-N^2)\hat{S}^E(k,l)}{2\mathrm{i}\beta k_+(k_+n+l_+m)-\mathrm{i}n\beta\left(K^2+K_s^2\right)/2+(\sigma+2r)K^2
K_s^2}\mathrm{d}k\mathrm{d}l\nonumber\\ &=&\upi(\sigma+r)N^2-\mathrm{i}\upi n\beta,\label{eq:dispersion_app}
\end{eqnarray}
where $K_s^2=(k+n)^2+(l+m)^2$. The corresponding dispersion relation on a periodic box, can be readily calculated by
simply substituting the integrals in (\ref{eq:dispersion_app}) by finite sums of integer wavenumbers. For a mirror symmetric
forcing obeying:
\begin{equation}
\hat\Xi(-k, l)=\hat\Xi(k, l),\label{eq:mirror}
\end{equation}
the eigenvalues $\sigma$ satisfy the symmetries (\ref{eq:symmet}). In
order to show this, we consider (\ref{eq:dispersion_app}) for $\sigma_{(-n, m)}$ and change the sign of $k$ in
the integral to obtain:
\begin{eqnarray}
& &\int_{-\infty}^{\infty}\int_{-\infty}^{\infty}\frac{(mk-nl)\left[nm(k_+^2-l_+^2)+(m^2-n^2)k_+l_+
\right]K^2(K^2-N^2)\hat{S}^E(-k,l)}{-2\mathrm{i}\beta k_+(k_+n+l_+m)+\mathrm{i}n\beta\left(K^2+K_s^2\right)/2+(\sigma_{(-n, m)}+2r)K^2
K_s^2}\mathrm{d}k\mathrm{d}l\nonumber\\ &=&\upi(\sigma_{(-n, m)}+r)N^2+\mathrm{i}\upi n\beta.\label{eq:symm_app1}
\end{eqnarray}
Taking the conjugate of (\ref{eq:symm_app1}) and using the mirror symmetry (\ref{eq:mirror}) yields
(\ref{eq:dispersion_app}) and therefore $\sigma_{(-n, m)}=\sigma_{(n, m)}^*$. Similarly, it can be readily
shown by considering (\ref{eq:dispersion_app}) for $\sigma_{(n, -m)}$ and changing the sign of $l$ in
the integral, that $\sigma_{(n, -m)}=\sigma_{(n, m)}$.

\end{document}